\documentclass[aps,prd,twocolumn,showpacs,amsmath,amssymb,nofootinbib]{revtex4}
\usepackage{bbm}
\usepackage{color}
\usepackage{graphicx}
\usepackage{times}

\newcommand{\Sw}{S_{\text{\tiny W}}}
\newcommand{\Seff}{S_{\text{\tiny eff}}}
\newcommand{\Spl}{S_{\text{p}}}
\newcommand{\Sj}{S^{j}}
\newcommand{\tr}{\text{tr}}
\newcommand{\dint}[1]{\mathfrak{D}#1}
\newcommand{\dU}{\mathfrak{D}U}
\newcommand{\dP}{\mathfrak{DP}}
\newcommand{\pol}{\mathfrak{P}}

\DeclareMathOperator{\sgn}{sgn}

\newcommand{\vev}[1]{\left\langle #1 \right\rangle}

\newcommand{\vc}[1]{\mbox{\boldmath$#1$}}
\newcommand{\svc}[1]{\mbox{\footnotesize\boldmath$#1$}}
\newcommand{\ssvc}[1]{\mbox{\scriptsize\boldmath$#1$}}
\newcommand{\Pol}[1]{\mathfrak{P}_{\svc{#1}}}

\newcommand{\Lx}{L_{\ssvc{x}}}
\newcommand{\Ly}{L_{\ssvc{y}}}

\newcommand{\Le}{L_{\mathrm{e}}}
\newcommand{\Lo}{L_{\mathrm{o}}}
\newcommand{\bLe}{\bar{L}_{\mathrm{e}}}
\newcommand{\bLo}{\bar{L}_{\mathrm{o}}}

\newcommand{\bra}{\langle}
\newcommand{\ket}{\rangle}

\newcommand{\crit}{\text{crit}}
\renewcommand{\Re}{\mbox{Re}\,}
\renewcommand{\Im}{\mbox{Im}\,}
\newcommand{\pad}[2]{\frac{\partial #1}{\partial #2}}

\begin{document}

\title{Phase Structure of $\boldsymbol{\mathbb{Z}(3)}$-Polyakov-Loop Models}

\author{Christian Wozar, Tobias Kaestner and Andreas Wipf}
\affiliation{Theoretisch-Physikalisches Institut,
Friedrich-Schiller-Universit{\"a}t Jena, Max-Wien-Platz 1, 07743
Jena, Germany}

\author{Thomas Heinzl}
\affiliation{School of Mathematics and Statistics, University of
Plymouth, Drake Circus, Plymouth, PL4 8AA, United Kingdom}

\author{Bal{\'a}zs Pozsgay}
\affiliation{Institute for Theoretical Physics, E{\"o}tv{\"o}s
Lor{\'a}nd University, P{\'a}zm{\'a}ny P{\'e}ter s{\'e}t{\'a}ny
1/A, 1117 Budapest, Hungary}

\begin{abstract}
We study effective lattice actions describing the Polyakov loop
dynamics originating from finite-temperature Yang-Mills theory.
Starting with a strong-coupling expansion the effective action is
obtained as a series of $\mathbb{Z}(3)$-invariant operators
involving higher and higher powers of the Polyakov loop, each with
its own coupling. Truncating to a subclass with two couplings we
perform a detailed analysis of the statistical mechanics involved.
To this end we employ a modified mean field approximation and
Monte Carlo simulations based on a novel cluster algorithm. We
find excellent agreement of both approaches concerning the phase
structure of the theories. The phase diagram exhibits both first
and second order transitions between symmetric, ferromagnetic and
anti-ferromagnetic phases with phase boundaries merging at three
tricritical points. The critical exponents $\nu$ and $\gamma$ at
the continuous transition between symmetric and anti-ferromagnetic
phases are the same as for the 3-state Potts model.
\end{abstract}


\maketitle

\section{Introduction}\label{sec:intro}

In two seminal papers Svetitsky and Yaffe have tentatively linked
the finite-temperature phase transitions in `hot' gauge theories
to the simpler order-disorder phase transitions of spin models
\cite{Yaffe:1982qf,Svetitsky:1982gs}. In the general case their
conjecture may be stated as follows: The effective theory
describing finite-temperature Yang-Mills theory with gauge group
$SU(N_C)$ in $d+1$ dimensions is a spin model in $d$ dimensions
with a global symmetry group given by the centre $\mathbb{Z}(N_C)$
of the gauge group. A somewhat stronger version of the conjecture
can be formulated if the phase transitions in questions are of
second order. In this case Yang-Mills theory and spin model fall
into the same universality class and the critical exponents
coincide. This has been convincingly demonstrated for $SU(2)$
\cite{Engels:1989fz, Engels:1992fs} in $d=3$.

However, continuous phase transitions in hot gauge theories are
not generic and hence the universality statement is almost empty
-- at least in 3+1 dimensions \cite{Holland:2003kg}. On the other
hand, the more general version of the statement has already been
used by Svetitsky and Yaffe to argue that the phase transition for
$SU(3)$ Yang-Mills theory must be first order for $d=3$ as there
is no $\mathbb{Z}(3)$  RG fixed point in this case. Since then
this has been firmly established by a number of lattice
calculations
\cite{Celik:1983wz,Kogut:1982rt,Gottlieb:1985ug,Brown:1988qe,Fukugita:1989yb,
Alves:1990yq}.

The conjecture implies that the effective theories may be
formulated as `Polyakov loop models'
\cite{Pisarski:2000eq,Pisarski:2001pe,Dumitru:2003hp}. For
$SU(N_C)$ this means that the macroscopic dynamical variables have
to reflect the complete gauge invariant information contained in
the (untraced) Polyakov loop, which in lattice notation is given
by
\begin{equation}
\label{eq:UTRPOLDEF}
 \Pol{x} [U] \equiv \prod_{t=1}^{N_{\tau}} U_{t,\ssvc{x};0} \; .
\end{equation}
This is a temporal holonomy winding around the compact Euclidean
time direction of extent $N_{\tau}$. For gauge groups with
nontrivial centre the traced Polyakov loop,
\begin{equation}
\label{eq:TRPOLDEF}
 \Lx \equiv  \tr_F \, \Pol{x}  \; ,
\end{equation}
with the trace being taken in the defining representation\footnote{We 
do not include a normalisation factor $1/N_C$ for the
ease of later notation.} serves as an order parameter for the
deconfinement phase transition. The phase transition goes along
with spontaneous breaking of the centre symmetry resulting from
\textit{non-periodic} gauge transformations under which
\begin{equation}
  \Lx \to z \Lx \; , \quad z \in \mathbb{Z}(N_C) \; .
\end{equation}
The deconfined broken-symmetry phase at sufficiently large Wilson
coupling $\beta > \beta_c$ is characterised by $\bra L \ket \ne 0$
(see \cite{Holland:2000uj} for a review).

Recently it has been found that gauge groups with trivial centre
may also lead to a deconfinement transition depending on the
\textit{size} of the gauge group
\cite{Borgs:1983yk,Holland:2003mc,Holland:2003jy}.

At this point the choice of dynamical variables needs to be
addressed. Under \textit{periodic} gauge transformations $g \in
SU(N_C)$ the holonomy (\ref{eq:UTRPOLDEF}) transforms as
\begin{equation}
 \Pol{x} \to g_{\svc{x}} \, \Pol{x} \, g^{-1}_{\svc{x}}  \; ,\label{eq:PGT}
\end{equation}
which leaves its eigenvalues and, as a consequence, its trace
(\ref{eq:TRPOLDEF}) invariant.  The eigenvalues are permuted
arbitrarily by gauge transformations corresponding to Weyl
reflections \cite{Ford:1998mq}. This invariance is taken into
account by constructing symmetric polynomials in the $N_C$
eigenvalues. From unimodularity the product of the eigenvalues is
$1$ and there are only $N_C-1$ independent polynomials, for
example the traces $\tr_F \Pol{x}^n$ for $1\leq n< N_C$. These in
turn are in one-to-one correspondence with the characters of the
$N_C -1$ fundamental representations (see below). Hence, for
$SU(2)$ $\Lx$ (which is real) is sufficient while $SU(3)$ requires
$\Lx$ and $\Lx^*$, the latter being a linear combination of
$\Lx^2$ and $\tr_F \, \Pol{x}^2$. Only for $N_C \ge 4$ traces of
higher powers of $\Pol{}$ are needed as independent dynamical
variables \cite{Meisinger:2001cq}.

In two recent papers \cite{Dittmann:2003qt,Heinzl:2005xv}
we have studied Polyakov loop models on the
lattice for the simplest non-Abelian gauge
group $SU(2)$. The models have been derived using strong-coupling
techniques at small Wilson coupling $\beta$ and a newly developed
inverse Monte Carlo (IMC) method which works for arbitrary values
of $\beta$. The latter method allows for a mapping of Yang-Mills
theory at a certain value of $\beta$ to any appropriately chosen
Polyakov loop model of the form
\begin{equation}
\label{eq:POLMODEL}
  S_{\mathrm{eff}} = \sum_{\bra \ssvc{x}\ssvc{y} \ket,IJ}
  \lambda_{IJ}(\beta)  \, \Big( \chi_I(\vc{x})  \chi_J (\vc{y}) +
  \text{h.c.} \Big)  \; ,
\end{equation}
where the summation is over nearest neighbours and group
representations $I$, $J$. The group character $\chi_I$
is the trace of the Polyakov loop in representation $I$,
\begin{equation}
\label{eq:CHARS}
 \chi_I (\vc{x}) \equiv \chi_I [\Pol{x}] \equiv \tr_I \, \Pol{x} \; .
\end{equation}
All characters $\chi_I$ are polynomials in the characters
corresponding to the fundamental representations. For $SU(2)$ the
simplest model is of Ising type,
\begin{equation}
\label{eq:SU2ISING}
 S_1 \equiv \lambda \sum_{\bra \ssvc{x}\ssvc{y} \ket} \Lx \Ly \; ,
\end{equation}
as first suggested  in \cite{Polonyi:1982wz} (see also
\cite{Ogilvie:1983ss,Billo:1996wv}).

The output of the IMC routines are the $\beta$-dependent effective
couplings $\lambda_{IJ}$.  Even without particular knowledge of
these one may study the models (\ref{eq:POLMODEL}) in their own
right as statistical field theories. The Svetitsky-Yaffe
conjecture may then be utilised to deduce information about the
Yang-Mills phase transition. For $SU(2)$ we have been able to show
that the mean-field analysis of the effective models yields a
surprisingly good agreement with the Monte Carlo analysis of both
the model itself and the underlying Yang-Mills dynamics. It should
be stressed at this point that the models (\ref{eq:POLMODEL}) are
not just simple scalar field theories with a linear target space
as all characters (\ref{eq:CHARS}) take values in a compact space
parametrised by the $N_C - 1$ fundamental characters. Hence, in
the lattice path integral each lattice site is endowed with the
reduced Haar measure on the gauge group rather than a Lebesgue
measure.

\begin{figure}[!ht]
\includegraphics{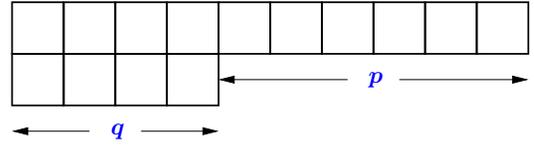}
\caption{\label{fig:PQ} Character labels and $SU(3)$ Young tableaux.}
\end{figure}

This paper presents our first steps to generalise the results of
\cite{Dittmann:2003qt,Heinzl:2005xv} to the gauge group $SU(3)$. We label the
characters by two integers $p$ and $q$ which count the numbers of
fundamental and conjugate representations (`quarks' and
`antiquarks' in the $SU(3)$ flavour language) required to
construct the representation $(p,q)$. Equivalently, these integers
characterise the horizontal extensions of $SU(3)$ Young tableaux
(see Fig.~\ref{fig:PQ}).
Under the $\mathbb{Z}(3)$ centre transformations the characters
transform according to the rule
\begin{multline}
\label{eq:CENTRE}
 \quad\chi_{pq} \to z_k^p z_k^{*q} \chi_{pq} \equiv z_k^{p-q} \chi_{pq}
 \; ,\\ \quad z_k \equiv \exp \left(\frac{2 \pi i}{3} k \right) \; ,
 \quad  k=0,1,2  \; ,\quad
\end{multline}
so that the most general centre-symmetric effective action with
nearest-neighbour interaction may be written as
\begin{multline}
\label{eq:SSU3}
  S_{\mathrm{eff}}[\chi] = \sum_{\substack{\bra \ssvc{x} \ssvc{y}\ket , \,
  pq, \, p'q' \\ p + p' = q + q' \, \mathrm{mod} \, 3}}
  \lambda_{pq, p'q'} \, \Big( \chi_{pq} (\vc{x})
  \, \chi_{p'q'} (\vc{y}) + \text{h.c.} \Big) \\
  + \sum_{\substack{\ssvc{x} , \, pq \\ p = q \, \mathrm{mod} \, 3 }}
  \lambda_{pq,00} \, \Big( \chi_{pq} (\vc{x}) + \text{h.c.} \Big) \; .
\end{multline}
This coincides with the ansatz suggested by Dumitru et al.\
\cite{Dumitru:2003hp}. The first sum in the effective action
consists of hopping terms involving monomials of the form $\Lx^m
\Ly^n$ or $\Lx^m \Ly^{*n}$ (and h.c.) while the second sum is a
`potential' term containing only powers $\Lx^n$ (and h.c.)
localised at single sites.

The remainder of the paper is organised as follows. In
Section~\ref{sec:SCE} we confirm the ansatz (\ref{eq:SSU3}) by
means of a strong coupling (small-$\beta$) expansion for
$S_{\mathrm{eff}}[\chi]$. For a restricted set of couplings (and
hence representations) we investigate the resulting
$\mathbb{Z}(3)$ models by minimising the classical action
(Section~\ref{sec:QCA}) followed by an improved mean field
analysis in Section~\ref{sec:MF}. In agreement with the
Svetitsky-Yaffe conjecture we find a first order phase transition
from the symmetric to a ferromagnetic phase. Our improved mean
field analysis already reveals an interesting phase structure with
four different phases: two distinct ferromagnetic phases, one
symmetric and one anti-ferromagnetic phase. Besides the first
order transitions we detect second order transitions from the
symmetric to an anti-ferromagnetic and to a ferromagnetic phase.
The continuous transition from the symmetric to the
anti-ferromagnetic phase is to be expected since the models reduce
to the $3$-state Potts model for Polyakov-loops having values in
the centre of the gauge group. Section~\ref{sec:MCSim} contains
the results of our extensive Monte Carlo simulations performed on
a Linux cluster where we have implemented the powerful package
jenLaTT. Similarly as for $SU(2)$ the mean field and numerical
results are in surprisingly good agreement. This is presumably due
to the existence of (three) tricritical points. Depending on the
order of the transition we localise the critical lines with 
either a Metropolis, a multicanonical or a modified cluster
algorithm. In addition we have checked that the critical exponents
$\nu$ and $\gamma$ at the second order transition to the
anti-ferromagnetic phase agree with those of the $3$-state Potts
model in $3$ dimensions. Finally, in Section~\ref{sec:SO} we wrap
up with discussion and conclusions.

\section{Strong-Coupling Expansion}
\label{sec:SCE}

In this section we briefly recapitulate the strong-coupling
(small-$\beta$) expansion for the $SU(3)$ Wilson action at finite
temperature \cite{Ogilvie:1983ss,Svetitsky:1985ye}. It is known
that the leading order result ($\beta^{N_{\tau}}$) stems from
ladder diagrams that wind around the temporal lattice extension
and corresponds to an Ising type model analogous to
(\ref{eq:SU2ISING}). By going beyond the leading order we will
encounter higher group representations/characters and hence have
an independent confirmation of the ansatz (\ref{eq:SSU3}) for the
effective action.

Our starting point is the standard Wilson action,
\begin{align}
\label{eq:wilsonactiondef}
 \Sw = \beta \sum_{\square} \left( 1-  \frac{1}{N_C} \Re \, \tr
 \, U_{\square} \right) \; ,
\end{align}
where the summation over plaquettes contains both temporal and
spatial links. The effective action $\Seff[\pol]$ is introduced as
usual by inserting an appropriate (group valued) delta function in
the path integral,
\begin{equation}
\begin{split}
  \mathcal{Z} &= \int \dU e^{-\Sw} \\
   &= \int \dP \int \dU \, \delta
  \left(\Pol{} , \prod_{\tau=0}^{N_\tau} U_{\tau,0} \right) \,
  e^{-\Sw[U]} \\
  & \equiv  \int \dP \, e^{-\Seff[\pol]} \; .
\end{split}
\end{equation}
While it is not known how to perform the final integration
analytically for the full Wilson action one can straightforwardly
integrate order by order in $\beta$. Thus we expand the Boltzmann
weight
\begin{equation}\label{eq:BOLTZMANN1}
e^{-\Sw} \equiv \sum_k \tilde{O}_{k} \beta^k \; ,
\end{equation}
and integrate separately over temporal and spatial links, $\dU =
\dint{U_t} \, \dint{U_s}$. Adopting temporal gauge we set all
temporal links equal to unity apart from the links
in the first timeslice  which according to
(\ref{eq:UTRPOLDEF}) may be identified with the Polyakov loop.
Integrating out all spatial link variables we obtain the partition
function
\begin{equation}
\begin{split}
 \mathcal{Z} & = \int \dP \, \exp \left( \log \sum_k O_k \, \beta^k
 \right) \\
 & \equiv \int \dP \, \exp \left( -\Seff[\pol] \right) \; ,
\end{split}
\end{equation}
where we have introduced the operators $O_k  \equiv \int
\dint{U_s} \tilde{O}_k$. The effective action may finally be
written as
\begin{align}
 \Seff = - \log \left(\sum_k O_k \, \beta^k \right) \equiv \sum_n S_n \beta^n \;
,
\end{align}
where the coefficients $S_n$ are related to the operators $O_k$
via the linked-cluster theorem. In the remainder of this section
we are going to determine the explicit form of the effective
operators $S_n$.

We first rewrite the Wilson-Boltzmann weight (\ref{eq:BOLTZMANN1})
as
\begin{align}
\label{eq:BOLTZMANN2}
  e^{-\Sw} \equiv \exp \left( {-\sum_{\square} \Spl} \right) = \prod_{\square}
e^{-\Spl} \; ,
\end{align}
and expand the single-plaquette contribution in terms of $SU(3)$ characters,
\begin{equation}
 e^{-\Spl} = \sum_{I} a_{I} (\beta) \, \chi_{I} (U_\text{p}) \; ,
\end{equation}
with $I \equiv (p,q)$ (see Fig.~\ref{fig:PQ}). All $\beta$
dependence now resides in the generalised Fourier coefficients
$a_{I}$ which accordingly may be further expanded,
\begin{align}
 e^{-\Spl} = \sum_{I,k}  a_{I}^k \, \beta^k \, \chi_{I} (U_\text{p} ) \; .
\end{align}
An explicit computation of the $a_{I}^k$ shows
that these vanish whenever the representation labels become
sufficiently large, namely if $|I| \equiv p+q > k$
\cite{Buss:2004}. This yields the important intermediate result
that to any given order $k$ in the strong-coupling expansion only
a \textit{finite} number of characters contributes,
\begin{align}
\label{eq:expansionSp}
  e^{-\Spl} = \sum_k \left( \sum_{|I|=0}^{k} a_I^k \,
  \chi_I(U_\text{p}) \right) \, \beta^k \; .
\end{align}
The integrations over the spatial links are standard group
integrals which can be found in the texts
\cite{Creutz:1984mg,Montvay:1994cy}. The upshot is that only
connected link arrangements (`polymers') wrapping around the
temporal extent of the lattice yield nonvanishing contributions.
The leading term is a ladder diagram consisting of $N_\tau$
plaquettes each of which contributes a factor of $\beta$ implying
a total contribution of $O(\beta^{N_\tau})$. The associated
operator is explicitly found to be
\begin{align}
 O_{N_\tau} \propto \chi_{10} (\Pol{x}) \chi_{01} (\pol_{\svc{x}+i}) +
\text{h.c.}\;.
\end{align}
A typical operator of order $\beta^{j N_\tau}$ is given by
\begin{equation}
\label{eq:actionbuildingblock}
 \Sj = \sum_{I,|I|=j} \sum_{x,i} C_I(\beta) \Big( \chi_I (\Pol{x})
 \chi_I^* (\pol_{\svc{x}+i}) + \text{h.c.}  \Big) \; ,
\end{equation}
in terms of which the Wilson-Boltzmann weight (\ref{eq:BOLTZMANN2}) becomes
\begin{equation}
\label{eq:effaction}
\begin{split}
 e^{-\Sw} & = \exp \left( - \ln \sum_n O_n \,\beta^n \right) \\
 & = c(\beta) +
\sum_{r=1
 \ldots k} \sum_{\substack{a_1 \ldots  a_r \\ a_1 +  \dotsb + a_r
 \leq k}} S^{a_1}  \dotsm S^{a_r} \; .
\end{split}
\end{equation}
Expanding this to next-to-leading order ($\beta^{2N_\tau}$) yields
the simple expression
\begin{equation}
\label{eq:S2N}
 \Seff = S^1 + S^2 + S^1S^1 + O(\beta^{3N\tau}) \; .
\end{equation}
Note, however, that some care has to be taken in interpreting
products such as $S^iS^j$ which by \eqref{eq:actionbuildingblock}
also contain disconnected pieces. Upon expanding the logarithm by
means of the linked-cluster theorem we are led to keep only
connected contributions of the form
\begin{multline}
 \quad S^i S^j \propto \sum_{\substack{I,|I|=i \\ J,|J|=j}} \sum_{\svc{x},k}
 \Big( \chi_I (\Pol{x})  \chi_I^* (\pol_{\svc{x}+k}) + \text{h.c.} \Big) \\
 \times \Big( \chi_J (\Pol{x}) \chi_J^* (\pol_{\svc{x}+k}) + \text{h.c.}
 \Big) \; .\quad
\end{multline}
Making use of the character reduction formula
\begin{equation}
 \chi_I (\vc{x}) \chi_J (\vc{x}) = \sum_K C_{IJ}^K \, \chi_K
(\vc{x}) \; , \quad C_{IJ}^K \in \mathbb{R} \; ,
\end{equation}
products of characters at the same site may be reduced to single
characters. As a consequence, the connected part of (\ref{eq:S2N})
takes the explicit form
\begin{equation}
\label{eq:effactionlong}
\begin{split}
 \Seff &\equiv \lambda_{10} \sum_{\svc{x},i} \Big( \chi_{10}(\Pol{x})
 \chi_{01}(\pol_{\svc{x}+i}) + \text{h.c.} \Big)
 \\
 &+ \lambda_{20} \sum_{\svc{x},i} \Big( \chi_{20}(\Pol{x})
\chi_{02}(\pol_{\svc{x}+i}) +
\text{h.c.} \Big) \\
 &+ \lambda_{11} \sum_{\svc{x},i}  \chi_{11}(\Pol{x})
\chi_{11}(\pol_{\svc{x}+i}) \\
 &+ \lambda_{21} \sum_{\svc{x},i} \Big( \chi_{20}(\Pol{x})
 \chi_{10}(\pol_{\svc{x}+i}) \\
 &\qquad + \chi_{10}(\Pol{x}) \chi_{20}(\pol_{\svc{x}+i}) +
\text{h.c.} \Big)
 \\
 &+ \rho_{1} \sum_{\svc{x}} \chi_{11}(\Pol{x}) + O(\beta^{3N_\tau})
 \; .
\end{split}
\end{equation}
For what follows it is useful to introduce the short-hand notation
\begin{equation}
\label{eq:effactionshort}
 \Seff = \lambda_{10} S_{10} + \lambda_{20} S_{20} + \lambda_{11}
 S_{11} + \lambda_{21} S_{21} + \rho_1 V_1 + O(\beta^{3N\tau}) \; ,
\end{equation}
with the obvious term-by-term identifications as compared to
(\ref{eq:effactionlong}). Our conventions are such that all
couplings in (\ref{eq:effactionlong}) and
(\ref{eq:effactionshort}) are real functions of $\beta$, the
single leading one being $\lambda_{10} = O(\beta^{N_\tau})$ (as
noted already in \cite{Ogilvie:1983ss}) while the subleading ones
are $O(\beta^{2N_\tau})$. It is straightforward to include
higher-order terms the number of which increases rapidly. At order
$\beta^{3N_\tau}$, for instance, there are already 11 terms so
that we refrain from going beyond next-to-leading order in
$\beta$.

For later purposes it is useful to express the operators appearing
in (\ref{eq:effactionshort}) in terms of the fundamental loops $L$
and $L^*$. Octet and sextet characters ($\chi_{11}$ and
$\chi_{20}$, respectively) are eliminated via the standard
reduction identities
\begin{eqnarray}
\label{eq:RED}
   3 \otimes 3^* = 8 \oplus 1\quad\hbox{and}\quad
   3 \otimes 3 = 6 \oplus 3^* \; ,
\end{eqnarray}
which are equivalent to the character relations (recall that
$L=\chi_{10}$)
\begin{eqnarray}
\label{eq:CHARRED}
  \chi_{11} =|L|^2 - 1 \quad\hbox{and}\quad
  \chi_{20} =L^2 - L^* \; .
\end{eqnarray}
Making use of the latter the different terms in
(\ref{eq:effactionshort}) become
\begin{align}
  S_{10} &= \sum_{\bra \ssvc{x} \ssvc{y} \ket} \left( \Lx \Ly^* +
  \text{h.c.} \right) \; , \label{eq:S10} \\
  S_{20} &= \sum_{\bra \ssvc{x} \ssvc{y} \ket} \left( \Lx^2 \Ly^{*2} - \Lx^2
  \Ly - \Lx^* \Ly^{*2} + \Lx^* \Ly + \text{h.c.} \right) \; , \\
  S_{11} &= \sum_{\bra \ssvc{x} \ssvc{y} \ket} \left( |\Lx|^2 |\Ly|^2 -
  |\Lx|^2 - |\Ly|^2 + 1 \right) \; , \\
  S_{21} &= \sum_{\bra \ssvc{x} \ssvc{y} \ket} \left( \Lx^2 \Ly - \Lx^*\Ly +
  \Ly^2 \Lx - \Lx \Ly^* + \text{h.c.} \right) \; , \label{eq:S21} \\
  V_1 &= \sum_{\ssvc{x}} \left( |\Lx|^2 - 1 \right) \; .
\end{align}
From these expressions it is obvious that each operator $S_{pq}$
is manifestly real. Under charge conjugation $\chi_{pq} \to
\chi_{pq}^* = \chi_{qp}$, hence $\Lx \to \Lx^*$, whereupon all
terms in $\Seff$ are charge conjugation invariant as required
\cite{Holland:2000uj}. Note that the orders included correspond to
terms that are quadratic, cubic and quartic in $L$ and/or $L^*$.
Higher powers will arise upon taking into account higher
representations. Thus, in this respect $S_{21}$ is somewhat
singled out being of only cubic order. This will become important
in a moment.

\section{Qualitative Classical Analysis}
\label{sec:QCA}

It has already been pointed out by Svetitsky and Yaffe
\cite{Yaffe:1982qf} that effective actions with $\mathbb{Z}(3)$
centre symmetry are closely related to the 3-state Potts model
which shows a first order phase transition from a symmetric to a
ferromagnetic phase \cite{KnakJensen:1979,Kogut:1982ss,Wu:1982ra}.
To make the relation manifest we restrict the Polyakov loop to the
centre elements $z_k$ introduced in (\ref{eq:CENTRE}). Setting
$\Pol{x} = z_k$ we find the general formula
\begin{equation}
 \chi_{pq} (z_k) = z_k^{p-q} \, d_{pq} \; ,
\end{equation}
where $d_{pq}$ denotes the dimension of the representation
$(p,q)$,
\begin{equation}
  d_{pq} = \frac{1}{2} (p+1)(q+1)(p+q+2) \; .
\end{equation}
Applying this to the effective action in
(\ref{eq:effactionshort}) we find, up to an additive constant,
\begin{equation}
  \Seff [z_k] = 
  \lambda \sum_ {\bra \ssvc{x} \ssvc{y} \ket}\cos \left(
  \frac{2\pi}{3}(k_{\ssvc{x}} - k_{\ssvc{y}}) \right) \;
  ,  \quad k_{\ssvc{x}}\in\{0,1,2\}\; ,\label{eq:pottsm}
\end{equation}
with effective coupling
\begin{equation}
  \lambda = 18\left( \lambda_{10} + 4\lambda_{20} + 4 \lambda_{21}
  \right) \; . \label{eq:PP}
\end{equation}
The action (\ref{eq:pottsm}) is precisely the one of the 3-state
Potts model \cite{Potts:1952,Wu:1982ra}. We thus expect that our
effective Polyakov loop models will have a phase structure
generalising the one of the 3-state Potts model. The latter is
known to have a ferromagnetic phase for large negative $\lambda$
and an anti-ferromagnetic phase for large positive $\lambda$
\cite{Janke:1996qb,Gottlob:1994ds}. As a preparation for the
discussion of later sections it is hence useful to obtain some
qualitative understanding of the phase structure of the Polyakov
loop models viewed as generalisations of the 3-state Potts model.
For the reason mentioned at the end of the previous section we
choose as a minimal generalisation the following effective action,
\begin{equation}
  \Seff \equiv \lambda_{10} S_{10} + \lambda_{21} S_{21} \; ,
\end{equation}
which in terms of the fundamental loops may be written explicitly
as
\begin{multline}
\label{eq:2CMODEL}
\quad  \Seff = (\lambda_{10} - 2 \lambda_{21}) \sum_{\bra \ssvc{x}
  \ssvc{y} \ket} \left( \Lx \Ly^* + \text{h.c.} \right) \\
  + \lambda_{21} \sum_{\bra \ssvc{x} \ssvc{y} \ket} \left( \Lx^2
  \Ly + \Ly^2 \Lx + \text{h.c.} \right) \; .\quad
\end{multline}
Note that there are also quadratic contributions stemming from
$S_{21}$. The action (\ref{eq:2CMODEL}) is manifestly
$\mathbb{Z}(3)$ centre symmetric under $\Lx \to z_k \Lx$.

\begin{figure}[!ht]
\includegraphics{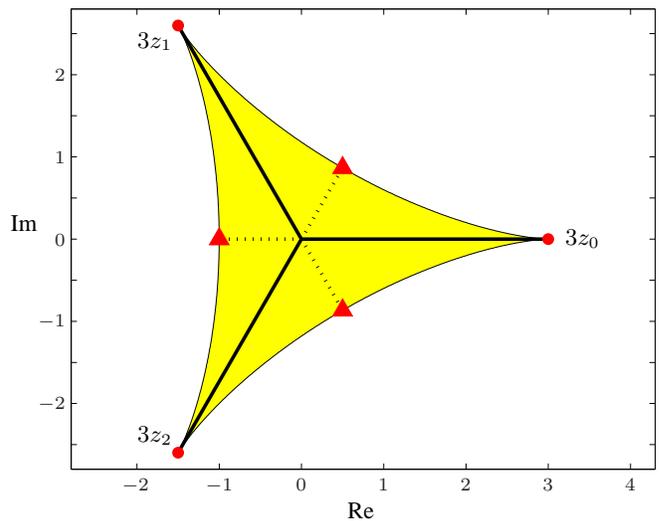}
\caption{\label{fig:DOMAIN} Target space of the Polyakov loop $L$
in the complex $L$-plane. The corners represent the three centre
elements. The intermediate points (denoted
anti-centre elements) will also become relevant for the discussion
of the phase structure.}
\end{figure}

It is important to realise that (\ref{eq:2CMODEL}) differs from
the standard lattice actions for scalar fields in several
respects. First, the field $\Lx$ is dimensionless, being the trace
of a unitary matrix. This allows for the presence of cubic hopping
terms connecting neighbouring sites. Even more important is the
fact that the target space of $\Lx$ is compact. Introducing the
eigenvalues of $\pol$ via
\begin{equation}
\label{eq:PDIAG}
  \pol_{\mathrm{diag}}  =  \text{diag} \Big( e^{i \phi_1},
  e^{i\phi_2}, e^{-i(\phi_1 + \phi_2)} \Big) \; ,
\end{equation}
and writing $L = L_1 + i L_2$ we find for the real and imaginary part
of $L$,
\begin{align}
  L_1 &= \cos \phi_1 + \cos \phi_2 + \cos (\phi_1 +
  \phi_2) \; , \label{eq:L1} \\
  L_2 &= \sin \phi_1 + \sin \phi_2 - \sin (\phi_1 +
  \phi_2) \; . \label{eq:L2}
\end{align}
The target space of $L$ may then be sketched in the complex
$L$-plane (see Fig.~\ref{fig:DOMAIN}). The boundary corresponds to
the points with $\phi_1 = \phi_2$,  the singular `corners' being
given by the three centre elements $\pol = z_k\mathbbm{1}$.
Let us try to get some first rough idea of the phase structure
associated with the two-coupling model (\ref{eq:2CMODEL}) in the
$\lambda_{10}$-$\lambda_{21}$ plane by looking at the extrema of
the classical action. If we vary the couplings these will trace
out a certain (possibly discontinuous) trajectory in the target
space given by the triangle of Fig.~\ref{fig:DOMAIN}.

\begin{figure}[!ht]
\includegraphics{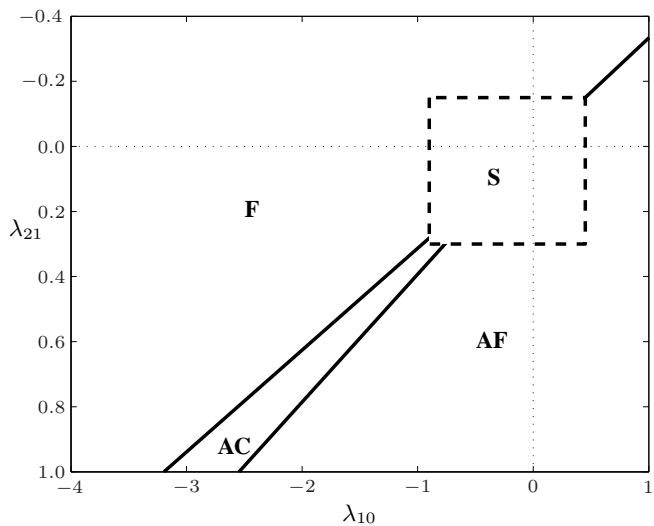}
\caption{\label{fig:CLPHASES} Qualitative prediction of the phase
diagram in the coupling constant plane for the effective action
(\protect\ref{eq:2CMODEL}). The ferromagnetic (F), anti-centre (AC)
and anti-ferromagnetic (AF) phases are obtained by looking for
classical minima. The symmetric, disordered phase (S) is located
where entropy is expected to dominate over energy.}
\end{figure}

As we argued earlier, for centre-valued Polyakov loops the
effective action (\ref{eq:2CMODEL}) reduces to the action of the
Potts model  (\ref{eq:pottsm}) with coupling
$\lambda=18(\lambda_{10}+4\lambda_{21})$. Thus we expect a
ferromagnetic phase (F) for large negative
$\lambda_{10}+4\lambda_{21}$ and an anti-ferromagnetic phase (AF)
for $\lambda_{10}+4\lambda_{21}$ large and positive. In a region
around the origin in the coupling plane entropy dominates energy
and we cannot expect to actually obtain the correct
phase-structure in this region by purely classical reasoning based
on minimising the energy. Qualitatively we expect a symmetric phase in
a neighbourhood of the origin. This is represented schematically
in Fig.~\ref{fig:CLPHASES} by the central rectangle.

In order to study the ordered phases (in particular AF) we divide
the lattice in two sub-lattices (denoted `even' and `odd') where the
Polyakov loop takes values $\Le$ and $\Lo$, respectively.
Two nearest neighbours belong to different sub-lattices.
The absolute minima of the classical action
\begin{multline}
\label{eq:smin}
\quad\Seff(\Le,\Lo)\propto (\lambda_{10}-2\lambda_{21})(\Le\Lo^*+\hbox{h.c.}) \\
+\lambda_{21}(\Lo^2\Le+\Le^2\Lo+\hbox{h.c.})\quad
\end{multline}
will then be located at certain values $\bLe$ and $\bLo$ of the
Polyakov loop which are identified with its `expectation values'.
We trust this reasoning as long as we are sufficiently far from
the origin of the coupling plane i.e. from the disordered,
entropy-dominated phase.

Any \textit{ferromagnetic} ordering will be characterised by a
minimum with $\bLe = \bLo =\bar{L} \neq 0$ while in an
\textit{anti-ferromagnetic} phase $\bLe \ne \bLo$. Quite
interestingly we find two distinct ferromagnetic phases, one for
which the Polyakov loop is near a centre element or $\bar{L}$ in
the vicinity of $3z_k$ and a different ferromagnetic phase with
$\bar{L}$ taking values near the intermediate points marked by
triangles in Fig.~\ref{fig:DOMAIN}. We call this an
\textit{anti-centre} phase (AC). We expect a phase transition line
separating the ferromagnetic and anti-ferromagnetic phases at
vanishing Potts-coupling $\lambda_{10}+4\lambda_{21}$. The
resulting qualitative phase diagram is depicted in
Fig.~\ref{fig:CLPHASES}.

To discuss the ferromagnetic phases it suffices to minimize
the action (\ref{eq:smin}) with $\Le=\Lo=L$, in which case
\begin{equation}\label{eq:sminn}
\Seff(L)\propto (\lambda_{10}-2\lambda_{21})\vert L\vert^2
+2\lambda_{21}(L^3+L^{*3}) \; .
\end{equation}
This can be done analytically. To localise the anti-ferromagnetic
phase we have calculated the absolute minima of (\ref{eq:smin}) on
the target space depicted in Fig.~\ref{fig:DOMAIN} numerically.
The combined analytical and numerical results are summarised as
follows. For negative $\lambda_{21}$ we have a transition
\begin{equation}
  \text{F} \xrightarrow{\lambda_{10} = -3 \lambda_{21}}
  \text{AF}\;
,
\end{equation}
while for positive $\lambda_{21}$ there is a richer
structure,
\begin{equation}
  \text{F}
\xrightarrow{\lambda_{10} = -3.1962 \lambda_{21}}
  \text{AC}
\xrightarrow{\lambda_{10} = - \frac{28}{11}
  \lambda_{21}} \text{AF} \; .
\end{equation}
\begin{figure}[!ht]
\includegraphics{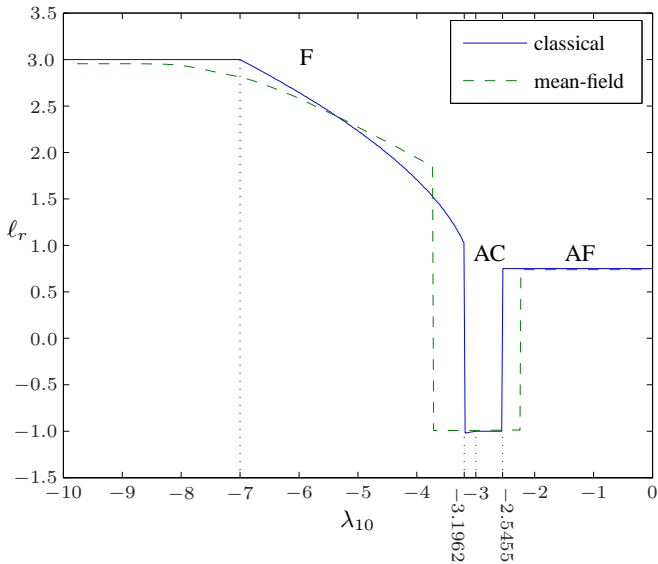}
\caption{\label{fig:classicAfer} Behaviour of the order parameter
$\ell_r$ defined in (\ref{eq:defprojector}) as a function of
$\lambda_{10}$ for fixed $\lambda_{21} = 1$.
For comparison we have added the result from the mean field
analysis to be developed in the next section.}
\end{figure}%
The behaviour of a suitably projected order parameter $\ell_r$
(the precise definition of which will only be needed later on) for
positive $\lambda_{21} = 1$ is shown in
Fig.~\ref{fig:classicAfer}. Upon inspection one notes that for
$\lambda_{10}$ sufficiently negative the system starts out with
the Polyakov loop at a centre element. Increasing $\lambda_{10}$
beyond $-7 \lambda_{21}$ the order parameter drops monotonically
until, at a critical coupling $\lambda_{10} \approx - 3.1962
\lambda_{21}$, there is a jump to the AC phase with $\bar{L}$ near
a anti-centre element. The jump of $\ell_r$ is due to a
centre-transformation and does not imply that the Polyakov loop
itself jumps. Indeed, $\bar L$ changes smoothly and arrives at an
anti-center element for $\lambda_{10}=-3\lambda_{21}$.  The system
stays there until $\lambda_{10} = -28 \lambda_{21}/11$, where it
jumps again, this time to the AF phase. As expected, we see no
symmetric phase in a purely classical analysis. Actually, for
$\lambda_{21}=1$ there is just no symmetric phase.

\section{Mean field approximation}
\label{sec:MF}

The next step of refinement to be presented in this section is a
mean field (MF) analysis of the effective action
(\ref{eq:2CMODEL}). This will serve as a basis for a comparison
with results from direct Monte Carlo simulations to be discussed
later on. Due to the peculiarities of the model as compared to
standard scalar field theories the application of the MF
approximation is not entirely straightforward. For the benefit of
the reader we will set the stage by giving a brief outline of the
necessary modifications. For further details the reader is
referred to our earlier paper \cite{Heinzl:2005xv}. To keep the
discussion sufficiently general we will first treat the effective
action (\ref{eq:effactionshort}) with five couplings focussing on
simpler examples later on.

We are interested in expectation values which are computed by
evaluating
integrals of the form
\begin{equation}
\label{eq:defobs}
\begin{split}
  \langle A \rangle &= \frac{1}{Z[0]} \int \mathcal{D}\pol \,
  e^{-\Seff[\pol]} \, A[\pol] \; , \\ \mathcal{D} \pol &\equiv
  \prod_{\ssvc{x}} d\mu(\pol_{\ssvc{x}}) \; ,
\end{split}
\end{equation}
which apparently extend over the whole group manifold employing
the Haar measure $d\mu(\Pol{x})$. However, due to the gauge
invariance of both action and measure the integrals can be reduced
to the coset space of conjugacy classes which we (somewhat
symbolically) denote by $\mathcal P_{\ssvc{x}}$. Hence we
integrate with the reduced Haar measure by replacing
\begin{equation}
  d\mu(\Pol{x}) \to d\mu_{\rm red}(\mathcal P_{\ssvc{x}}) \; .
\end{equation}
Thus, \eqref{eq:defobs} is equivalent to
\begin{equation}
\label{eq:polyakovmeasure}
\begin{split}
  \langle A \rangle &= \frac{1}{Z[0]} \int \mathcal{DP} \,
  e^{-\Seff [\mathcal P]} \, A[\mathcal P] \; ,\\
  \mathcal{DP} &= \prod_{\ssvc{x}} d\mu_{\rm red}(\mathcal P_{\ssvc{x}})\;.
\end{split}
\end{equation}
The probability measure $\mathcal{DP} \, \exp(-\Seff)/Z[0]$ is
characterized as the unique solution to the variational problem
\begin{equation}
\label{eq:definfimum}
  \inf_p \langle \Seff + \log p \rangle_p = - \log Z[0] = -W[0] \;.
\end{equation}
The expectation value on the left-hand side has to be taken with
respect to the integration measure $ \prod_{\ssvc{x}} d\mu_{\rm red}(\mathcal
P_{\ssvc{x}})\,
p[\mathcal{P}]$ with $p[\mathcal{P}]$
denoting the probability density of $\mathcal{P}$. From this point
of view the MF approximation is nothing else but the restriction of the
permissible densities to product form,
\begin{equation}
p[\mathcal P] \to p_{\mathrm{mf}}[\mathcal P] \equiv
\prod_{\ssvc{x}} p_{\ssvc{x}} (\mathcal P_{\ssvc{x}})\; .
\end{equation}
Expectation values can now be simply computed site by site via
factorisation,
\begin{equation}
  \langle \chi_I(\mathcal{P}_{\ssvc{x}} )
  \chi_J(\mathcal{P}_{\ssvc{y}}) \rangle \to \langle
  \chi_I(\mathcal{P}_{\ssvc{x}}) \rangle_{\ssvc{x}}
  \langle\chi_J(\mathcal{P}_{\ssvc{y}}) \rangle_{\ssvc{y}} \; .
\end{equation}
Our goal is to compute the effective potential as
function of the mean characters
$u_{\textrm{mf}}=u_{\textrm{mf}}(\bar\chi_{I})$.
For that purpose we solve the variational problem 
(\ref{eq:definfimum}) on the space of product 
measures with fixed expectation values of the
characters. This is done by introducing appropriate 
Lagrange multipliers $j_{I}$. 
For ferromagnetic systems one may
assume that the weight functions $p_{\ssvc{x}}$ at each site are
identical, $p_{\ssvc{x}}=p$. This assumption corresponds to a
translationally invariant ground state.
According to the
discussion of the previous section we expect anti-ferromagnetic
phases and hence we must refine our choice for $p_{\ssvc{x}}$. We
therefore introduce different weight functions on the even and odd
sub-lattices, respectively,
\begin{equation}
  p_{\ssvc{x}}(\mathcal P_{\ssvc{x}}) =
  \begin{cases}
  p_{\mathrm{e}}(\mathcal P_{\ssvc{x}}) \; &:\; \text{sgn} (\vc{x}) =  1 \; ,\\
  p_{\mathrm{o}}(\mathcal P_{\ssvc{x}}) \; &:\; \text{sgn} (\vc{x}) = -1 \; ,
  \end{cases}
\end{equation}
defining the sign of a lattice point as
\begin{equation}
  \text{sgn} (\vc{x}) \equiv (-1)^{^{\textstyle \sum_i x_i}} \; .
\end{equation}
As a consequence, expectation values of characters will
subsequently have two values depending on the sub-lattice where
they are evaluated,
\begin{equation}
\langle \chi_I \rangle_{\ssvc{x}} =
\begin{cases}
\bar \chi_{I, \mathrm{e}} \; &:\; \text{sgn} (\vc{x}) =  1 \; ,\\
\bar \chi_{I, \mathrm{o}} \; &:\; \text{sgn} (\vc{x}) = -1 \; .
\end{cases}
\end{equation}
The sources are taken constant as well when restricted to the even
and odd sub-lattices, $j_{I,\ssvc{x}} \equiv j_{I,\mathrm{e}}$ or
$j_{I,\mathrm{o}}$, respectively. Like the characters the sources
are complex.

The action \eqref{eq:effaction} couples only nearest-neighbour
sites so that its expectation value entering (\ref{eq:definfimum})
may be written as
\begin{multline}
\quad  \langle \Seff \rangle = V d \, \Big\{ \lambda_{10} \,  ( \bar
  \chi_{10,\mathrm{o}} \bar \chi_{01,\mathrm{e}} + \text{h.c.}) + \dotsb
  \Big\} \\
  + \frac{V}{2} \rho_1 \, \left(\bar\chi_{11,\mathrm{o}} +
  \bar\chi_{11,\mathrm{e}}\right) \; ,\quad
\end{multline}
with $V = N^d$ denoting the lattice volume in $d$ spatial
dimensions. The logarithm in (\ref{eq:definfimum}) decomposes as
\begin{equation}
  \langle \log p \rangle = \frac{V}{2} \big(\langle \log
  p_{\mathrm{o}} \rangle_{\mathrm{o}} + \langle \log
  p_{\mathrm{e}} \rangle_{\mathrm{e}} \big) \; .
\end{equation}
It is convenient to drop the common volume factor $V$ and consider
densities instead. The variation of \eqref{eq:definfimum} finally
yields the weight function for the even sub-lattice,
\begin{multline}
\label{eq:solweightfunc}
\quad  p_{\mathrm{e}} (\mathcal P) = \frac{1}{z(\vc{j}_{\mathrm{e}}, \,
  \vc{j}^*_{\mathrm{e}})} \exp \big\{ - \rho_1 V_1 (\mathcal P) \\
  + \vc{j}_{\mathrm{e}} \cdot \vc{\chi}(\mathcal P) +
  \vc{j}_{\mathrm{e}}^*\cdot \vc{\chi}^*(\mathcal P)\big \} \; .\quad
\end{multline}
and a completely analogous expression for the odd sub-lattice.
Here we have introduced $\vc{j}\cdot \vc{\chi}$ as a short-hand
for $\sum_I j_I \,\chi_I$ and the single-site partition function
\begin{multline}
\quad z(\vc{j} , \, \vc{j}^*) \equiv \int d \mu_{\rm red}(\mathcal P) \exp
\Big\{ -\rho_1
  V_1 (\mathcal P) \\
  + \vc{j} \cdot \vc{\chi} (\mathcal P) + \vc{j}^* \cdot
  \vc{\chi}^*(\mathcal P) \Big\} \; .\quad
\end{multline}
The sources $\vc{j}_{\mathrm{o,e}}$ are eliminated by inverting
the relations
\begin{equation}
  \bar{\chi}_I (\vc{j}, \vc{j}^*) = \pad{}{j_I} \, w(\vc{j}, \vc{j}^*) \; , \quad
  \bar{\chi}_I^* (\vc{j}, \vc{j}^*) = \pad{}{j_I^*} \, w(\vc{j}, \vc{j}^*) \; ,
\end{equation}
to be evaluated separately on both sublattices.  The Schwinger
function $w(\pmb j, \pmb j^*)$ is defined as usual,
\begin{equation}
  \label{eq:schwingerfunc}
  w(\pmb j, \pmb j^*) \equiv \log z(\pmb j , \, \pmb j^*) \; .
\end{equation}
Introducing the Legendre transform of \eqref{eq:schwingerfunc}
according to
\begin{equation}
  \gamma_0 (\bar{\vc{\chi}}, \bar{\vc{\chi}}^*) \equiv \inf_{\pmb
  j, \, \pmb j^*} \Big \{ \pmb j \cdot \bar{\vc{\chi}} + {\pmb
  j}^* \cdot \bar{\vc{\chi}^*} - w (\pmb j, \pmb j^*) \Big\}
\end{equation}
the solution of \eqref{eq:definfimum} is finally obtained as the
MF potential (density) as a function of even and odd mean fields,
\begin{multline}
\label{eq:distrsolution}
  u_{\mathrm{mf}}(\bar{\vc{\chi}}_{\mathrm{e}} ,
  \bar{\vc{\chi}}_{\mathrm{e}}^* , \bar{\vc{\chi}}_{\mathrm{o}},
  \bar{\vc{\chi}}_{\mathrm{o}}^*)
  = d \, \Big\{ \lambda_{10} (\bar{\chi}_{10,\mathrm{o}}
\bar{\chi}_{01,\mathrm{e}}
  + \text{h.c.}) + \ldots \Big\} \\
  + \frac{1}{2} \rho_1 (\bar\chi_{11,\mathrm{o}} + \bar\chi_{11,\mathrm{e}})
  + \frac{1}{2} \gamma_0 (\bar{\vc{\chi}}_{\mathrm{o}} ,
\bar{\vc{\chi}}^*_{\mathrm{o}})
  + \frac{1}{2} \gamma_0 (\bar{\vc{\chi}}_{\mathrm{e}} ,
\bar{\vc{\chi}}^*_{\mathrm{e}})\, .
\end{multline}
From this expression one can easily derive relations between the
sources $\pmb j$ and mean characters $\vc{\chi}$. For instance, by
varying $u_{\mathrm{mf}}$ with respect to $\bar{\chi}_{10,
\mathrm{o}}$ we obtain
\begin{equation}
  \label{eq:critpoint}
  0 = d \, \big( \lambda_{10} \bar{\chi}_{01,
  \mathrm{e}} + \lambda_{21} \bar{\chi}_{20, \mathrm{e}} \big) +
  \frac{1}{2}j_{10, \mathrm{o}} \; ,
\end{equation}
where we have used that the current is given as
\begin{equation}
  j_{10,\mathrm{o}} = \pad{}{\bar\chi_{10, \mathrm{o}}} \,
  \gamma_0 (\bar{\vc{\chi}}_{\mathrm{o}} ,
  \bar{\vc{\chi}}_{\mathrm{o}}^* ) \; .
\end{equation}
The first term in (\ref{eq:distrsolution}) derives directly from
the effective action (\ref{eq:effactionshort}) and hence contains
four couplings to which the potential coupling $\rho_1$ has to be
added. A complete MF analysis of this system becomes very awkward.
In what follows, we therefore specialise to effective actions with
only one or two couplings.

\subsection{One Coupling}

The action (\ref{eq:S10}) defines what we call the `minimal model'
with one coupling only,
\begin{equation}
\label{eq:SGP}
  \Seff \equiv \lambda S_{10} = \lambda \sum_{\bra \ssvc{x}
  \ssvc{y} \ket} \left( \Lx \Ly^* + \text{h.c.} \right) \; .
\end{equation}
Identifying $(\bar{\chi}_{10})_{\mathrm{e,o}} \equiv
L_{\mathrm{e,o}}$ and inserting (\ref{eq:SGP}) the MF potential
\eqref{eq:distrsolution} simplifies to
\begin{multline}
\label{eq:simplesolution_2}
\quad  u_{\mathrm{mf}} (\Le, \Le^*, \Lo, \Lo^*) = d \, \lambda \,
  (\Le \Lo^* + \Lo \Le^*) \\
  + \frac{1}{2} \gamma_0 (\Le, \Le^*) + \frac{1}{2} \gamma_0
  (\Lo, \Lo^*) \; .\quad 
\end{multline}
It is useful to define the generic MF potential
\begin{equation}
  v_{\mathrm{mf}} (L,L^*) \equiv d \, \lambda \, |L|^2 + \gamma_0 (L,L^*) \; ,
\end{equation}
so that (\ref{eq:simplesolution_2}) can be rewritten as
\begin{multline}
\label{eq:simplesolution}
\quad   u_{\mathrm{mf}}(\Le, \Le^*, \Lo, \Lo^*) = - \frac{d \lambda}{2}
  |\Lo - \Le|^2 \\
  + \frac{1}{2} v_{\mathrm{mf}} (\Le, \Le^*) +
  \frac{1}{2} v_{\mathrm{mf}}(\Lo, \Lo^*) \; .\quad 
\end{multline}
This expression clearly shows that for negative $\lambda$
configurations with $\Lo=\Le$ are favoured
making the distinction between even and odd sub-lattices obsolete.
The remaining unique expectation value,
\begin{equation}
\label{eq:defL}
  L = \frac{1}{V} \sum_x L_x = \frac{1}{2}\left(\Le + \Lo\right)\;,
\end{equation}
thus serves as an order parameter for the ferromagnetic phase
transition. On the other hand, if $\lambda > 0$ $\Le$ and $\Lo$
will cease to be positively correlated and we expect an
anti-ferromagnetic phase transition. In this case, a reasonable
order parameter is given by \cite{Wu:1982ra,Wang:1990}
\begin{equation}
\label{eq:defM}
  M \equiv \frac{1}{V}  \sum_{\ssvc{x}} \Lx \sgn(\vc{x})
   = \frac{1}{2}\left( \Le - \Lo \right) \; .
\end{equation}
Let us also introduce the absolute values of the order parameters,
henceforth denoted by
\begin{equation}
\label{eq:deflbarmbar}
 \ell  \equiv |L|\quad \text{ and }\quad \mathit m \equiv |M| \; .
\end{equation}
If we assume for the moment that the occurring phase transitions
are second order the corresponding MF critical couplings can be
computed analytically as we are going to demonstrate next. The
consistency condition \eqref{eq:critpoint} reduces for  the case
at hand to
\begin{equation}
\label{eq:gap}
  0 = d \lambda \Lo + \frac{1}{2} j_{\mathrm{e}} \quad  \text{and}
  \quad  0 = d \lambda \Le + \frac{1}{2}j_{\mathrm{o}} \; .
\end{equation}
As both $\Le$ and $\Lo$ may be assumed small near the critical
coupling $\lambda_c$ the corresponding sources will also be small.
Thus, we may expand
\begin{equation}
  L = \frac{\int d \mu_{\rm red} \, \exp\{ j \chi_{10}
  + \text{h.c.} \} \chi_{10}}{\int d\mu_{\rm red} \, \exp\{
  j \, \chi_{10} + \text{h.c.} \}} =
  j + O(j^2) \; ,
\end{equation}
which again holds separately on each sublattice, $\Le \simeq
j_{\mathrm{e}}$ and $\Lo \simeq j_{\mathrm{o}}$. Plugging this
into \eqref{eq:gap} yields $(2d\lambda)^2=1$ and hence the
critical couplings
\begin{equation}
  \lambda_\pm = \pm \frac{1}{2d} \; .
\end{equation}

For arbitrary order parameters we have solved the gap equations
(\ref{eq:gap}) numerically. The result is depicted in the
following figure. The transition S-AF is second order, the one to
the phase F first order. Thus, the anti-ferromagnetic transition
must be at $\lambda_+ = 1/2d$. The first order ferromagnetic
transition is slightly above $-1/2d$.
\begin{figure}[!ht]
\includegraphics{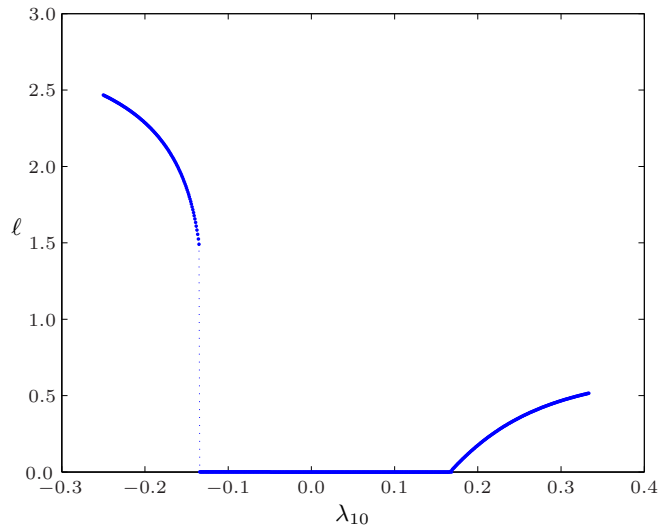}
\caption{\label{fig:meanScan} Mean field results for the minimal
model with one coupling. The first order ferromagnetic transition
S-F is at $\lambda_-=-0.13433$ and the second order transition
S-AF at $\lambda_+=0.166667$.}
\end{figure}%
Our MF analysis thus confirms the qualitative results from the
preceding section that already in the simplest model
there is both a ferromagnetic and an
anti-ferromagnetic transition. This is qualitatively
consistent with the phase diagram Fig.~\ref{fig:CLPHASES}
restricted to the horizontal axis.

\subsection{Two Couplings}\label{sec:MF.numerical}

Even for the simple model of the previous subsection no explicit
expression is known for $z_0(\pmb{j},\pmb{j}^*)$ as the $SU(3)$
group integrals cannot be evaluated in closed form (a fact well
known from strong-coupling expansions, see e.g.\
\cite{Montvay:1994cy}). Things naturally become worse if
additional couplings are turned on. Already for two couplings
i.e.\ the action (\ref{eq:2CMODEL}), the only way to proceed is by
means of numerical methods.

In order to obtain the MF version of the phase diagram
Fig.~\ref{fig:CLPHASES} we have employed the following algorithm:

\begin{enumerate}
\item At the extremal points of \eqref{eq:distrsolution} all sources
$\pmb j_{\mathrm{o,e}}$ occurring in \eqref{eq:solweightfunc} can
be eliminated in favour of the expectation values
$\bar{\boldsymbol\chi}_{\mathrm{o,e}}$ as in \eqref{eq:critpoint}.
Since the character target space is compact it can be easily
discretised defining measures $p_{\mathrm{e}}$ and
$p_{\mathrm{o}}$ at each point. Using these measures expectation
values $\langle\boldsymbol\chi\rangle_{\mathrm{o,e}}$ can now be
computed which in general will differ from
$\bar{\boldsymbol\chi}_{\mathrm{o,e}}$. Hence, we first look for
local minima of
\begin{equation}
 \sigma(\bar{\boldsymbol\chi}_{\mathrm{o}},\bar{\boldsymbol\chi}_{\mathrm{e}})=
 \Vert \langle\boldsymbol\chi\rangle_{\mathrm{o}}
 - \bar{\boldsymbol\chi}_{\mathrm{o}} \Vert +
 \Vert
 \langle\boldsymbol\chi\rangle_{\mathrm{e}} - \bar{\boldsymbol\chi}_{\mathrm{e}}
 \Vert,
\end{equation}
with norm $\Vert\pmb v\Vert \equiv \sum_i |v_i|$. These minima,
however, do not correspond to exact solutions yet but rather serve
as the starting points of a recursion.
\item
We now solve the equation
$\sigma(\bar{\boldsymbol\chi}_{\mathrm{o}},\bar{\boldsymbol\chi}_{\mathrm{e}}
)=0$ for $\bar{\boldsymbol\chi}_{\mathrm{o}}$ and
$\bar{\boldsymbol\chi}_{\mathrm{e}}$ by Newton iteration using the
local minima of the previous step as initial input. In this way we
end up with multiple solution vectors
$\bar{\boldsymbol\chi}_{\mathrm{o,e}}$ each extremising
\eqref{eq:distrsolution}.

\item
The solution vector with minimal $u_{\mathrm{mf}}$ contains
the desired ground state expectation values.
\end{enumerate}

With the help of this algorithm we are able to compute the expectation values of
$L$ \eqref{eq:defL}  and $M$ \eqref{eq:defM}.

We conclude this section with some remarks concerning our choice
of the sources $\pmb j$. Allowing them to be complex yields an
8-dimensional parameter space for the observables of model
(\ref{eq:2CMODEL}) with two characters, $\chi_{10}$ and
$\chi_{21}$. For this large parameter space it would actually be
simpler to perform a high-precision Monte Carlo simulation than to
find a good approximation for the global minima of
\eqref{eq:distrsolution}. For this reason, we have chosen real
sources as an input to our MF approximation. We expect this to be
a very good approximation as long as the peak of the probability
distribution for $\chi_{10}$ is concentrated near the real axis in
Fig.~\ref{fig:DOMAIN}.

Comparing with the classical analysis we note that our real-source
assumption is justified for all couplings which are located away
from the boundary between the phases F and AC. The F-AC
transition should be second order according to the analysis of
Section \ref{sec:QCA} while the MF approximation with real sources
predicts a first order transition (see
Fig.~\ref{fig:classicAfer}). The contradiction will be finally
resolved by Monte Carlo simulations showing that near the F-AC
transition the peaks of the probability distribution for
$\chi_{10}$ are off the real axis. Apart from this fairly small
region in parameter space we find a remarkable agreement between
the MF results and the Monte Carlo results of the next section.
This provides the ultimate justification for our simplifying
choice of real sources in the MF approximation.

\section{Monte Carlo Simulation}
\label{sec:MCSim}

The previous two sections have provided us with a good deal of
information on the phase structure of the effective model in the
most interesting regions of parameter space. Based on this we have
performed a large number of Monte Carlo simulations to
quantitatively check the MF predictions and obtain a precise picture
of the critical behaviour. As before, to avoid excessive complexity,
we have concentrated on the action (\ref{eq:2CMODEL}) with two
couplings $\lambda_{10}$ and $\lambda_{21}$.

At the beginning of Section~\ref{sec:MF} we have already noted
that both the action and the reduced Haar measure depend only on
the conjugacy class of the (untraced) Polyakov loop. Hence, one
can choose the basic field variables either as the traces in the
fundamental representations ($L$, $L^*$) and powers thereof or as
a suitable parametrisation in terms of the eigenvalues as
introduced in (\ref{eq:PDIAG}). It turns out that for a numerical
treatment the latter proves to be more appropriate and so we use
(\ref{eq:PDIAG}) to represent the conjugacy class according to
\begin{equation}
\label{eq:defangulardecomp}
\begin{split}
  [\pol_x] &\equiv \mathcal P_x (\boldsymbol \Phi) = \text{diag}
  \big( e^{i\phi_1}  , e^{i\phi_2}, e^{-i(\phi_1+\phi_2)}
  \big) \; ,\\
  \quad \boldsymbol\Phi &\equiv (\phi_{1},\phi_{2}) \; .
\end{split}
\end{equation}
Here, the following restrictions should be imposed such that the
angular coordinates cover each class only once,
\begin{equation}
\label{eq:fundDom}
  0 \leq \phi_1, \phi_2, \quad \phi_1 < \phi_2, \quad
  \phi_2 < (-\phi_1 - \phi_2) \bmod 2\pi \; .
\end{equation}
These restrictions are somewhat awkward to implement in a
simulation code. It is much more convenient to let $\boldsymbol
\Phi$ take values in the full square $[0,2\pi) \times [0,2\pi)$
which covers the fundamental domain given by (\ref{eq:fundDom})
six times. Due to the residual gauge symmetry of the system it is
clear that expectation values will be unaffected by this
over-counting.

In the coordinates (\ref{eq:defangulardecomp}) the reduced Haar
measure becomes
\begin{multline}
\label{eq:redHaar}
  d\mu_{\mathrm{red}}(\boldsymbol \Phi) = \frac{8}{3\pi^2} \sin^2 \left(
  \frac{\phi_1 - \phi_2}{2} \right)\\
  \times \sin^2 \left( \phi_1 + \frac{\phi_2}{2} \right)
  \sin^2 \left( \phi_2 + \frac{\phi_1}{2} \right) d\phi_1
  d\phi_2 \; ,
\end{multline}
where the normalization is such that the measure integrates to unity
over the square $[0,2\pi) \times [0,2\pi)$. All characters can be
expressed in terms of $\phi_1$ and $\phi_2$. Using (\ref{eq:redHaar}) the
full measure \eqref{eq:polyakovmeasure} may be straightforwardly
expressed in terms of the angular coordinates as
\begin{equation}
\label{eq:HaarPhi}
  \mathcal{DP} \, e^{-\Seff[\mathcal P]} = \prod_x d\mu_{\mathrm{red}}
  (\boldsymbol \Phi_x) e^{-\Seff[\boldsymbol \Phi]} \; .
\end{equation}

The focus of our numerical studies has been the phase diagram in the
$\lambda_{10}-\lambda_{21}$ plane. There, we have scanned through
the region $[-0.25,0.33] \times[-0.22\ldots 0.16]$ with a resolution
of $71\times 46$ points, which were in total $3266$ different Monte
Carlo simulations. Before we present our results a few words on our
numerical techniques are in order.

\subsection{Algorithms}

For the investigation of phase transitions, in particular their
order, histogram methods are widely used and accepted. This
approach, however, requires large statistics and thus tends to
consume a lot of computer time. In addition, we are interested in
a fairly large range of coupling constants. For these reasons, the
updating algorithm for the Polyakov loop models has to be fast and
versatile. It turns out that the standard Metropolis algorithm
favourably matches both requirements if we aim at an accuracy of
about $5\%-10\%$. On the other hand, because of the highly
nontrivial probability measure involved, a heat bath algorithm
does not seem applicable or, in any case, would be too time
consuming. In addition, its local nature should not yield any
enhancement of statistics near a first order phase transition. We
have thus refrained from implementing a heat bath update scheme
but rather decided to optimise the Metropolis algorithm as
described in the following two paragraphs.

\subsubsection{Multicanonical algorithm}

When a system undergoes a first order phase transition, the
histogram associated with the order parameter will generically
display a multi-peak structure. Depending on the total volume of the
system the peaks can be very pronounced. In other words, the
configuration space is decomposed into distinct sectors between
which local algorithms can hardly mediate. One way to overcome the
resultant failure in sampling the total configuration space is to
make use of the multicanonical algorithm, see e.g.\
\cite{Berg:1998nj}.

The crucial improvement step consists in the replacement of the
measure used in (\ref{eq:HaarPhi}) according to
\begin{equation}
  d\mu_{\mathrm{red}} \, \exp(-S) \to d \mu_{\mathrm{red}} \, \exp(-S) \,
\eta(\ell)
  \; .
\end{equation}
The new improved measure on the right-hand side has a weight $\eta
= \eta (\ell)$ which depends on the (modulus of the) order
parameter. One chooses the particular form
\begin{equation}
 \eta(\ell) \equiv \rho^{-1}(\ell) \; ,
\end{equation}
where $\rho(\ell)$ denotes the probability density of the order
parameter. This choice leads to an enhancement of configurations
that would otherwise be suppressed and thus allows for a much
improved ergodic behaviour of the algorithm.

\begin{figure}
\includegraphics{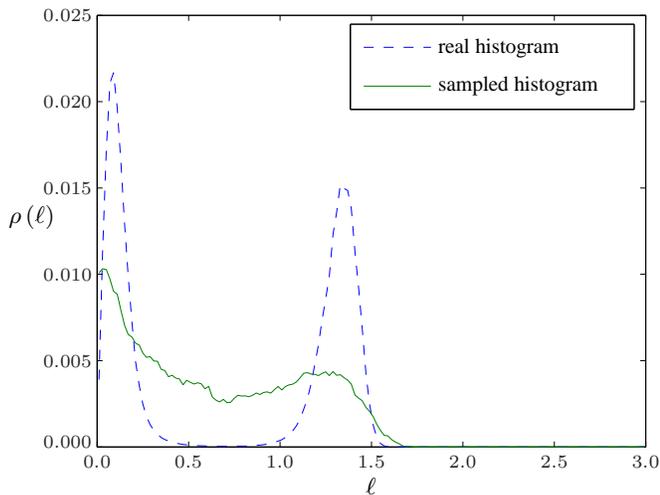}
\caption{\label{fig:multiCanHist}Histogram of $\ell$ with
$\lambda_{10}=-0.13721$ on a $10^3$-lattice sampled against the
histogram for $N=9$.}
\end{figure}

The effect due to the altered measure is illustrated in
Fig.~\ref{fig:multiCanHist} where two typical distributions are
plotted. In the original distribution one clearly recognises two
well-separated peaks. Any local algorithm will fail to sample such a
distribution properly once it is trapped at one of its peaks. The
distribution actually used `closes the gap' enabling transitions
between different peak regions during the simulation. In the end, of
course, one has to correct for the change in the measure by
reweighting with $\eta^{-1}$,
\begin{equation}
  \langle Q \rangle = \frac{\langle Q \, \eta^{-1}(\ell)
  \rangle_{\mathrm{mult.}}}{\langle \eta^{-1}(\ell) \rangle_{\mathrm{mult.}}} =
  \frac{\langle Q \, \rho(\ell) \rangle_{\mathrm{mult.}}}{\langle
  \rho(\ell)\rangle_{\mathrm{mult.}}} \; ,
\end{equation}
where we have denoted expectation values taken with respect to the
modified measure with a subscript `mult'.

\begin{figure}[!ht]
\includegraphics{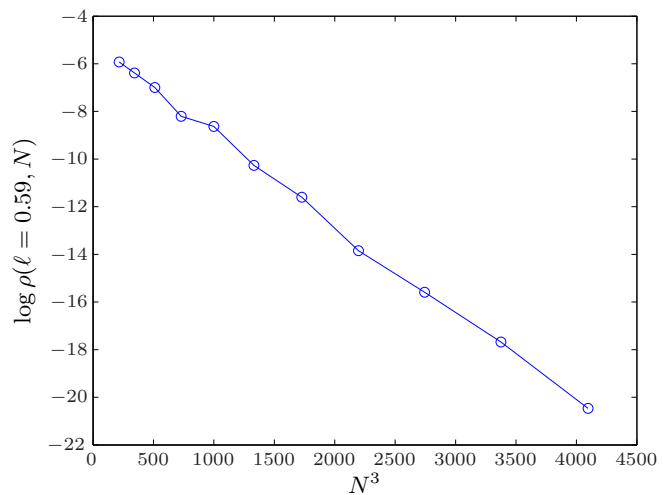}
\caption{\label{fig:multCanLog}Plot of $\log\rho(\ell,N)$ against
volume $V = N^3$ for the minimal model with $\lambda_{10} =
-0.13721$ confirming the scaling behaviour \eqref{eq:scalingRho}.
The coupling is very close to critical.}
\end{figure}

A slight problem with this approach, however, still has to be
overcome. One actually needs right at the beginning what one set
out to compute originally, namely the distribution $\rho$. A
practicable strategy is e.g.\ the following.  From a small lattice
volume, say $V_0$, where peaks are usually less pronounced, one
obtains an approximate distribution function $\rho_0(\ell)$ which
is then used on a slightly bigger lattice, say of volume $V_1$.
This simple trick can be further refined if on the larger lattice
one first computes $\rho_1(\ell)$ using $\rho_0$ and subsequently
repeats all measurements employing $\rho_1$. In practice, this
procedure is iterated several times to make larger and larger
lattices available. Going beyond a volume of $V=12^3$ requires
additional knowledge of the scaling behaviour of $\rho(\ell)$.
Fig.~\ref{fig:multCanLog} shows that, to a very good
approximation, the scaling depends linearly on the volume, $V =
N^3$,
\begin{equation}
\label{eq:scalingRho}
  \log \rho (\ell,N) \approx A(\ell) + C(\ell) \,  N^3 \; .
\end{equation}
In summary, the multicanonical algorithm yields substantial
improvements compared to the standard Metropolis algorithm and
allows for very accurate simulations. Lattices with volumes up to
$V=20^3$ could thus be studied near the first order phase
transition. However, as the implementation is very involved we
applied it only to the minimal model ($\lambda_{21}=0$). In
principle, the generalisation to incorporating further couplings is
straightforward, being merely a matter of having sufficient
computing time available.

\subsubsection{Cluster algorithm}

The multicanonical algorithm is best suited for studying first
order transitions. This is no longer true for second order
transitions where it is outperformed by cluster algorithms -- at
least if there is one available.  For our purposes, none of the
algorithms on the market could be immediately put to use.  We
therefore decided to modify the well-known Wolff algorithm
\cite{Wolff:1988uh}, an extension of the Swendsen-Wang
\cite{Swendsen:1987ce} algorithm originally proposed for discrete
spin systems.

Before actually describing our modifications let us briefly recall
the main idea of \cite{Wolff:1988uh}. Suppose the action (and the
measure) are invariant under a certain global symmetry which acts on
the fields according to
\begin{equation}
  \vc{\Phi}_{\svc x} \to \vc{\Phi}_{\svc x}' = R \, \vc{\Phi}_{\svc x} \; .
\end{equation}
Typically, the symmetry operator $R$ will depend on some
parameters which we collectively denote by $\omega$ (continuous or
discrete) so that $ R =  R(\omega)$. It is important to note that
the operator $R$ has to be idempotent, $R^2 = \text{id}$, in order
to ensure detailed balance. The algorithm then works according to
the following list.

\begin{enumerate}
\item Fix some parameter $\omega_0$ and hence some transformation $
R_0 \equiv R(\omega_0)$. Randomly choose a lattice point $\vc{x}$
and apply the symmetry transformation
\begin{equation}
  \vc{\Phi}_{\svc x}' = R_0 \, \vc{\Phi}_{\svc x} \; ,
\end{equation}
which may be viewed as flipping the field variable $\vc{\Phi}_{\svc x}$.
The point $\vc{x}$ is checked and added to the cluster.
\item\label{it:flipnn}
Repeat the following for all unchecked neighbours $\vc{y}$ of
$\vc{x}$:

Let $\tilde S = \tilde S(\vc{\Phi}_{\svc x} , \vc{\Phi}_{\svc y})$ denote the
contribution to the action from the link $\langle \vc{x} \vc{y} \rangle$
and compute
\begin{equation}
  \Delta \tilde S \equiv \tilde S(\vc{\Phi}_{\svc x}', R_0 \, \vc{\Phi}_{\svc
y}) -
  \tilde{S}(\vc{\Phi}_{\svc x}', \vc{\Phi}_{\svc y}) \; ,
\end{equation}
which may be rewritten as
\begin{equation}
\label{eq:deltaTildeS}
  \Delta \tilde S = \tilde S(\vc{\Phi}_{\svc x}, \vc{\Phi}_{\svc y}) -
  \tilde S(\vc{\Phi}_{\svc x}', \vc{\Phi}_{\svc y}) \; ,
\end{equation}
since $\tilde S$ is already invariant under $R$.
The decision to add $\vc y$ to the cluster
is subject to an accept/reject step so that the probability to flip
$\vc{\Phi}_{\svc y}$ becomes
\begin{equation}
  p =
  \begin{cases}
    0\; & :\; \Delta \tilde S>0\; , \\
    1 - e^{\Delta \tilde S}\; & :\; \Delta \tilde S<0 \; .
  \end{cases}
\end{equation}
If $\boldsymbol \Phi_{\svc y}$ is flipped,
check the point  $\vc{y}$.

\item Go back to Step~(\ref{it:flipnn}) for all sites added to the
cluster in the previous step.
\end{enumerate}
Note that \eqref{eq:deltaTildeS} measures whether it is advantageous to
flip $\vc{\Phi}_{\svc y}$ once $\vc{\Phi}_{\svc x}$ has
been flipped.
From the very construction of the algorithm it should be already
clear that the clusters will increase with the correlation length of
the system. In this way one suppresses the phenomenon of critical
slowing down observed with local algorithms. This makes the cluster
algorithm particularly suited for the study of second order phase
transitions.

\begin{figure}[!ht]
\includegraphics{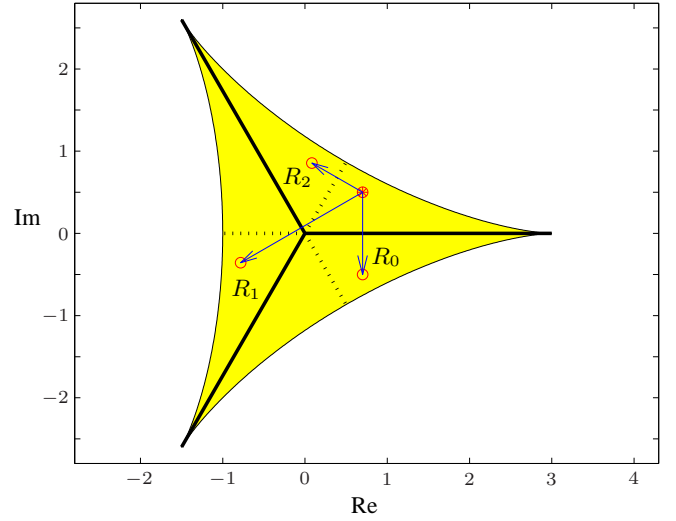}
\caption{\label{fig:reflections}Illustration of the
$\mathbb{Z}(3)$ reflections $R_i$ used in our modified cluster
algorithm.}
\end{figure}

The first step in adapting the cluster algorithm to our needs is
to find suitable symmetries of the action \eqref{eq:effaction}.
Being a Polyakov loop model the symmetry in question is the
discrete $\mathbb{Z}(3)$ symmetry, $L \to z_k L$. In addition, the
action has to be real which implies constraints on the way complex
conjugation acts. From these symmetries one can construct three
operators $R_i$, $i = 0,1,2$, acting on the Polyakov loop
according to
\begin{equation}
\label{eq:flipaction}
  R_i \, L = (z_i \, L)^* \; , \quad z_i \in \mathbb{Z}(3)
  \; .
\end{equation}
As required the $R_i$ square to unity. Furthermore, it is easy to
see that both the operators $S_{pq}$ appearing in
\eqref{eq:effactionshort} and the domain of the Polyakov loop $L$
are left invariant by the action of $R_i$. The latter is
illustrated in Fig.~\ref{fig:reflections} for a particular value
of $L$.
For the actual algorithm the transformations \eqref{eq:flipaction}
are not immediately applicable since the simulation is based on
the angular variables $\vc\Phi$ introduced in
(\ref{eq:defangulardecomp}). However, it is just a matter of a
little algebra to show that the $R_i$ act on the angles
$\vc\Phi$ via
\begin{equation}
  (\phi_1, \phi_2)
  \mapsto
  \begin{cases}
    (2\pi - \phi_1, 2\pi - \phi_2)  & \; \text{under $R_0$} \; , \\
    \left(\frac{4\pi}{3}-\phi_1,\frac{4\pi}{3}-\phi_2\right) \bmod
    2\pi & \; \text{under $R_1$} \; ,\\
    \left(\frac{2\pi}{3}-\phi_1,\frac{2\pi}{3}-\phi_2\right) \bmod
    2\pi & \; \text{under $R_2$} \; . \\
  \end{cases}
\end{equation}
Whereas in the original cluster algorithm flipping along randomly
chosen lattice sites is sufficient to guarantee ergodicity this is
no longer true in the present case. Hence, we have to augment the
update scheme by standard Metropolis sweeps to make the algorithm
ergodic. For one Monte Carlo step our cluster algorithm finally
can be summarised as follows:
\begin{enumerate}
\item Choose a random number $N_M$ between $0$ and $V = N^3$.
\item Do $N_M$ standard Metropolis sweeps at randomly drawn lattice points.
\item For a suitable fixed number $N_{\mathrm{cl}}$ repeat the
steps for building a cluster as described above.
\item Do $V - N_M$ additional Metropolis sweeps, again at randomly chosen
lattice sites.
\end{enumerate}
From several test runs we have found that the number
$N_{\mathrm{cl}}$ should be chosen such that total number of flipped
sites after performing step (3) is approximately half the number of
all lattice sites. Thus, if $|\mathcal C|$ denotes the typical size
of a cluster, the following equation should hold (at least
approximately),
\begin{equation}
 N_{\mathrm{cl}} = \frac{V}{2 |\mathcal C|} \; .
\end{equation}
One of the most interesting questions of course is the gain in
performance compared to e.g.\ the standard Metropolis algorithm.
We have found that the autocorrelation time for the order
parameter $\tau_{\ell}$ is independent of both the lattice extent (at least in
the range $N=8\ldots 28$)
and the coupling constant (if close to the second order phase
transition) implying a dynamical critical exponent of $z=0$.
Moreover, for an optimal choice of $N_{\mathrm{cl}}$ the
autocorrelation time is of order unity. As a result, our cluster
algorithm outperforms the Metropolis algorithm even for small
lattices. On the largest lattices we have considered the cluster
algorithm reduces autocorrelation times by two orders of magnitude
as compared to Metropolis updating. This improvement comes at the
cost of a slightly increased complexity, specifically a factor of
1.5 in computing time which clearly is negligible. On the other
hand, in line with our expectations, no significant improvement
has been found near the first order transition.

\subsection{Results}

We present the results of our Monte Carlo simulations in the same
two steps as for the MF approximation. Hence, we first report on
the minimal model (\ref{eq:SGP}) and switch on $\lambda_{21}$
later on. With only $\lambda_{10}$ different from zero it is
reasonably cheap  to perform highly accurate measurements so
that a precise quantitative comparison with the 3-state Potts
model is possible.

For the model (\ref{eq:2CMODEL}) with two couplings we determine
the phase diagram in the $\lambda_{10}$--$\lambda_{21}$ plane and
compare with our expectations as laid out in the previous two
sections. We conclude with a careful study of the nature of the
phase transitions, in particular their continuity properties.

\subsubsection{Minimal model (one coupling)}

\begin{figure}[!ht]
\includegraphics{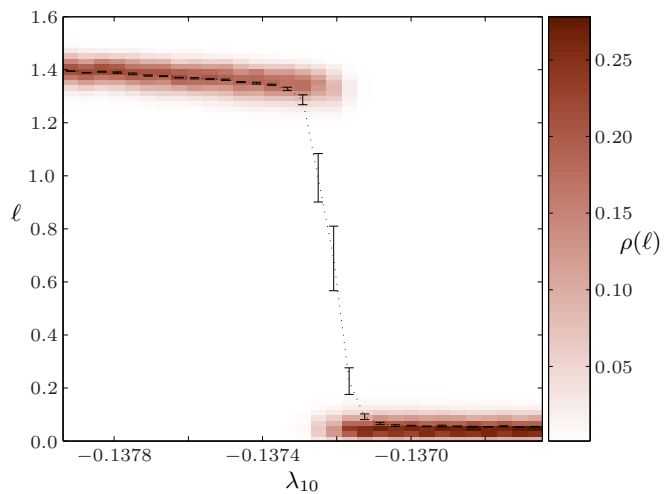}
\caption{\label{fig:ferroPtMc16} Ferromagnetic phase transition
computed with the multicanonical algorithm on a $16^3$-lattice.
The expectation value of $\ell$ is plotted against its probability
distribution given by the area shaded in grey. The latter clearly
shows the correct discontinuous behaviour. Note that, since we
measure the modulus, statistical fluctuations manifest themselves
in a (small) positive value of the order parameter even in the
symmetric phase. }
\end{figure}

To determine the ferromagnetic phase transition for $\lambda_{10}
< 0$ we have used standard techniques which need no further
explanation. It suffices to note that in order to study the first
order (S-F) transition we performed $10^6$ sweeps on a
$16^3$-lattice in the multicanonical ensemble for each value of
the coupling constant. This led to highly accurate statistics
until we approached the close vicinity of the critical coupling
itself. In this regime our statistics is restricted to $10^2$
independent samples due to large correlation times of the order of
$10^4$ sweeps. As Fig.~\ref{fig:ferroPtMc16} shows this is
sufficient to demonstrate that the transition is first order.

\begin{figure}[!ht]
\includegraphics{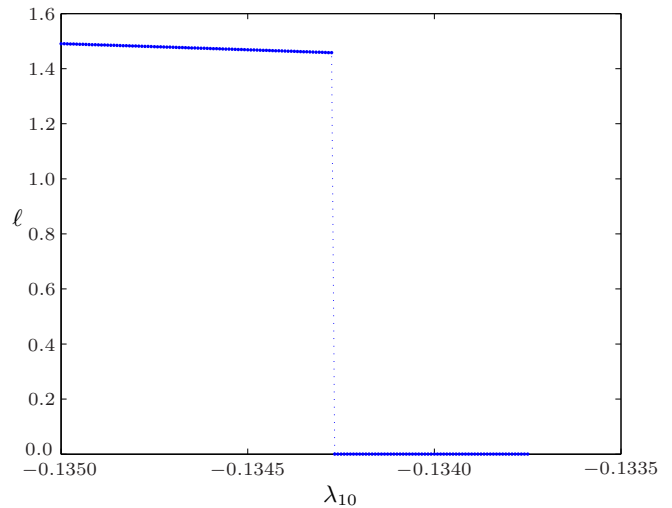}
\caption{\label{fig:ferroMF} Phase transition of
Fig.~\protect\ref{fig:ferroPtMc16} when
calculated within MF approximation.}
\end{figure}

In Fig.~\ref{fig:ferroMF} we compare our Monte Carlo result for
$\ell (\lambda_{10})$ with its MF approximation. The figure
basically zooms into that part of Fig.~\ref{fig:meanScan} where
the first order ferromagnetic transition is located. Again, the
first order nature of the transition is corroborated.

In addition to the coupling dependence of the order parameter we
have determined both the critical coupling and the discontinuity
$\Delta \ell$ at $\lambda_{10,\mathrm{crit}}$. The results are given in
Table~\ref{tab:crit1} together with the MF prediction. Again we
find a surprisingly good \textit{quantitative} agreement between
simulations and MF approximation. Moreover, if we consider the
ratio of the two critical couplings, say $\lambda_{10,\crit;
\mathrm{F}} / \lambda_{10,\crit;\mathrm{AF}}$, we are able to
compare this with results of the 3-state Potts model. The actual
figures turn out to be fairly close, namely $-0.6904$ for the
minimally coupled Polyakov model and $-0.6750$ for the 3-state
Potts model \cite{Janke:1996qb,Gottlob:1994ds}.
\begin{table}[!ht]
\caption{\label{tab:crit1} Critical couplings for the S-F and S-AF
phase transitions and jump $\Delta \ell$. For the
first order transition (S-F) there is excellent agreement between
MF and Monte Carlo data. Even the values for $\Delta \ell$
agree within 10\%. For the second order transition (S-AF) the
critical couplings agree within approximately 20\%.}
\begin{ruledtabular}
\begin{tabular}{lccc}
method &$\lambda_{10,\crit}$ (S-F) &$\Delta \ell$ &$\lambda_{10,\crit}$ (S-AF)\\
\hline
 Monte Carlo & $-0.13721(5)$ & $1.33(2)$ & $0.19875(5)$ \\
 Mean Field & $-0.13433(1)$ & $1.46(1)$ & $0.16667(1)$ \\
\end{tabular}
\end{ruledtabular}
\end{table}

\begin{figure}[!ht]
\includegraphics{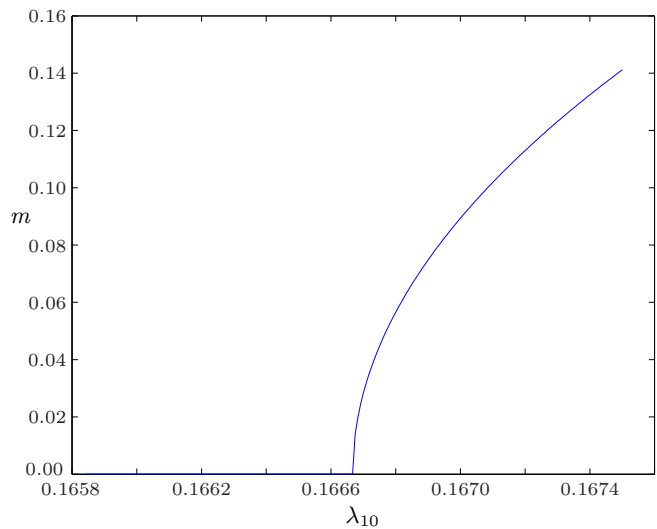}
\caption{\label{fig:aferrMF} Expectation value of $\mathit m$ near
the anti-ferromagnetic phase transition computed via MF
approximation.}
\end{figure}
\begin{figure}[!ht]
\includegraphics{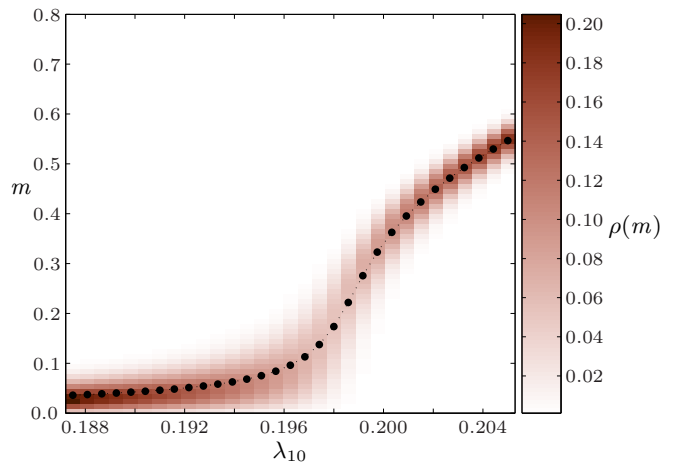}
\caption{\label{fig:aferrMcs} Expectation value and probability
distribution of $\mathit m$ near the anti-ferromagnetic phase
transition obtained from Monte Carlo simulations. To identify clear
signals we have chosen a large lattice with $V=28^3$  and evaluated
$5 \times 10^5$ sweeps. In contrast to Fig.~\ref{fig:ferroPtMc16} no
discontinuity is observed. Again the expectation value of the
symmetric phase is biased since we measure the modulus of the order
parameter.}
\end{figure}%

Table~\ref{tab:crit1} also displays the (positive) critical coupling
for the second order AF transition. This has been analysed with our
modified cluster algorithm by performing a total number of $2\times
10^6$ sweeps. With these large statistics at hand it is also
possible to determine some of the critical exponents thus probing
the universality properties of the model. To do so we have employed
standard renormalisation group techniques following
\cite{Gottlob:1993zd}. In particular, we consider the Binder
cumulant $U$ and susceptibility $\chi$ given by
\begin{align}
  U &= 1 - \frac{\vev{\mathit m^4}}{3\vev{\mathit m^2}^2} \; , \\
  \chi &= N^3\vev{\mathit m^2} \; ,
\end{align}
with $\mathit m$ as defined in \eqref{eq:defM} and $N$ denoting the
spatial extent of the lattice as before.

\begin{figure}[!ht]
\includegraphics{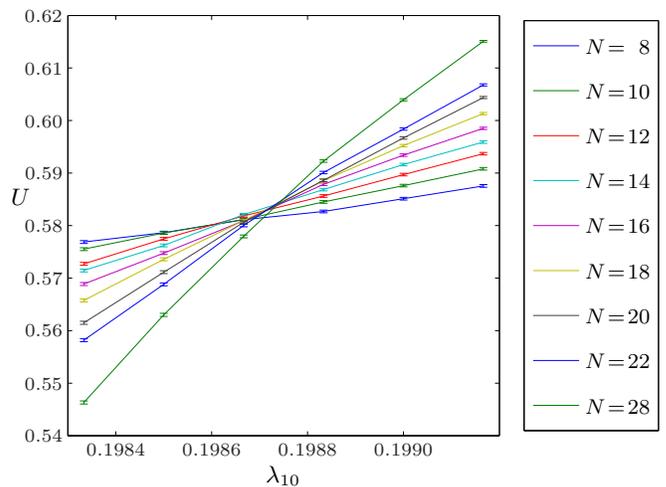}
\caption{\label{fig:binderCumulant} Binder cumulants $U$ for different lattice
sizes as a
function of the coupling $\lambda_{10}$. The critical coupling
$\lambda_{10,\crit}$ is determined via the intersection
point.}
\end{figure}

The Binder cumulant $U = U(N,\lambda_{10})$ is constructed such
that it becomes independent of $N$ close to the critical point.
Hence, the latter is  rather precisely determined as the point
where the graphs of $U$ (plotted for different $N$) intersect.
This behaviour is nicely exhibited in
Fig.\ref{fig:binderCumulant}.

From the standard relations at criticality,
\begin{gather}
\label{eq:critExpDef}
  \chi(\lambda_{10,\crit}) \propto N^{\gamma/\nu} \; ,\\
  \label{eq:critExpDef2}
  \left.\frac{\partial U(N,\lambda_{10})}{\partial
  \lambda_{10}}\right|_{\lambda_{10}=\lambda_{10,\crit}} \propto
  N^{1/\nu}  \; ,
\end{gather}
we have finally computed the critical exponents $\gamma$ and $\nu$
which are listed in Table~\ref{tab:critexpo}.

\begin{table}[!ht]
\caption{\label{tab:critexpo} Critical exponents for the
second order AF transition of the minimal model.}
\begin{ruledtabular}
 \begin{tabular}{ccc}
  exponent & 3-state Potts  \cite{Gottlob:1994ds} & minimal Polyakov\\ \hline
  $\nu$ & $0.664(4)$ & $0.68(2)$ \\
  $\gamma/\nu$ & $1.973(9)$ & $1.96(2)$ \\
 \end{tabular}
 \end{ruledtabular}
\end{table}

The uncertainty in $\lambda_{10,\crit}$ quoted in
Table~\ref{tab:crit1} is mainly due to the fact that the different
cumulants do not precisely meet in a single intersection point
(cf.\ again Fig.~\ref{fig:binderCumulant}). The error in the
critical exponents is estimated from a least square fit to the
logarithm of (\ref{eq:critExpDef}) and \eqref{eq:critExpDef2}. To
cross-check our results we have measured the expectation value of
$\mathit m$ and its probability distribution $\rho(\mathit m)$ in
analogy with the ferromagnetic transition already discussed. As in
the former case, the expectation value of the order parameter
alone does not suffice to decide on the order of the transition.
However, the probability distribution shown in
Fig.~\ref{fig:aferrMcs} is quite different from the one in
Fig.~\ref{fig:ferroPtMc16}. No discontinuous behaviour is observed
now which provides further (numerical) evidence for a second order
phase transition. Equally important, the critical coupling
obtained is compatible with the results presented in Table
\ref{tab:crit1}. For the sake of completeness
Fig.~\ref{fig:aferrMF} shows the MF prediction for the transition
S-AF.

Comparing with earlier results on the 3-state Potts model
\cite{Wang:1990} we draw the important conclusion that the minimal
Polyakov loop model is in the same universality class.

\subsubsection{The phase diagram for two couplings}

\begin{figure}[!ht]
\includegraphics{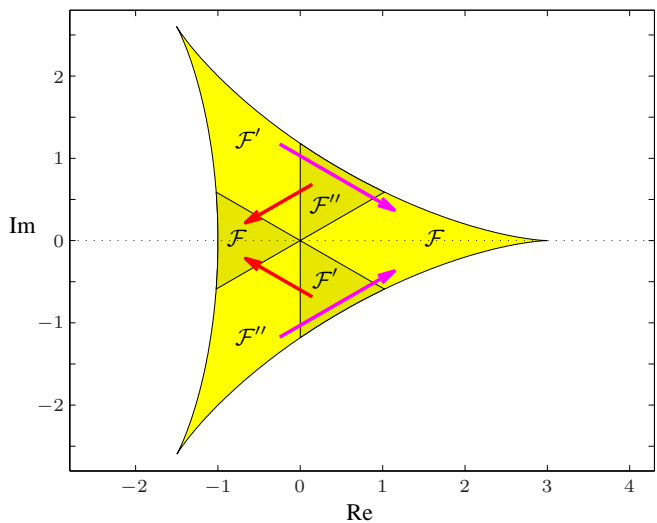}
\caption{\label{fig:rotatedObs} Fundamental domain $\mathcal{F}$
of the order parameter $L$ obtained by identifying $\mathbb{Z}(3)$
copies according to the depicted arrows.}
\end{figure}

Having discussed the minimal model at length let us continue by
switching on the second coupling $\lambda_{21}$ in order to
analyse the phase diagram in the coupling constant plane. This
requires a suitably chosen `indicator' to distinguish the (at
least) four phases (S, F, AC and AF) we expect in accordance with
our MF analysis of Section~\ref{sec:MF}. While $\ell$ and
$\mathit{m}$ clearly are order parameters for the minimal model
they are numerically less suited for the model with two couplings.
It turns out advantageous to construct a new observable denoted
$\ell_r$ which may be obtained from $\ell$ by the following
procedure. We first divide the domain of $L$ into six distinct
parts as shown in Fig.~\ref{fig:rotatedObs}. The light-shaded
region represents the preferred locus of the Polyakov loop in the
ferromagnetic phase F, whereas the dark-shaded region corresponds
to the anti-centre ferromagnetic phase AC. To eliminate the
(numerically superfluous) $\mathbb{Z}(3)$ symmetry the first step
in our projection is to identify the regions as indicated by the
arrows in Fig.~\ref{fig:rotatedObs}. In this way we end up with a
fundamental domain $\mathcal{F}$ for the $\mathbb{Z}(3)$ symmetry
centred along the real axis. Every $L$ is mapped into
$\mathcal{F}$ by a centre transformation. To finally obtain a real
observable we project the transformed $L$ onto the real axis. This
projection results in the variable $\ell_r$ the sign of which
clearly distinguishes between the two ferromagnetic phases.
$\ell_r<0$ indicates the AC phase, $\ell_r>0$ the ferromagnetic
phase F, while $\ell=0$ in the symmetric phase S. Mathematically,
the projection of $L$ to $\ell_r$ is given by
\begin{equation}\label{eq:defprojector}
 \ell_r =
 \begin{cases}
  \phantom{-\frac{1}{2}}\Re L \; & :\;
  L\in \mathcal{F}\; , \\
  -\frac{1}{2}\Re L + \frac{\sqrt{3}}{2}\Im L \; & :\;
  L\in \mathcal{F'}\; ,  \\
  -\frac{1}{2}\Re L-\frac{\sqrt{3}}{2}\Im L \; & :\;
  L\in \mathcal{F''}\; .
  \end{cases}
\end{equation}
To detect the AF phase we simultaneously measure $\mathit{m}$ so
that we can finally discriminate between all possible phases.

\begin{figure*}[!p]
\includegraphics{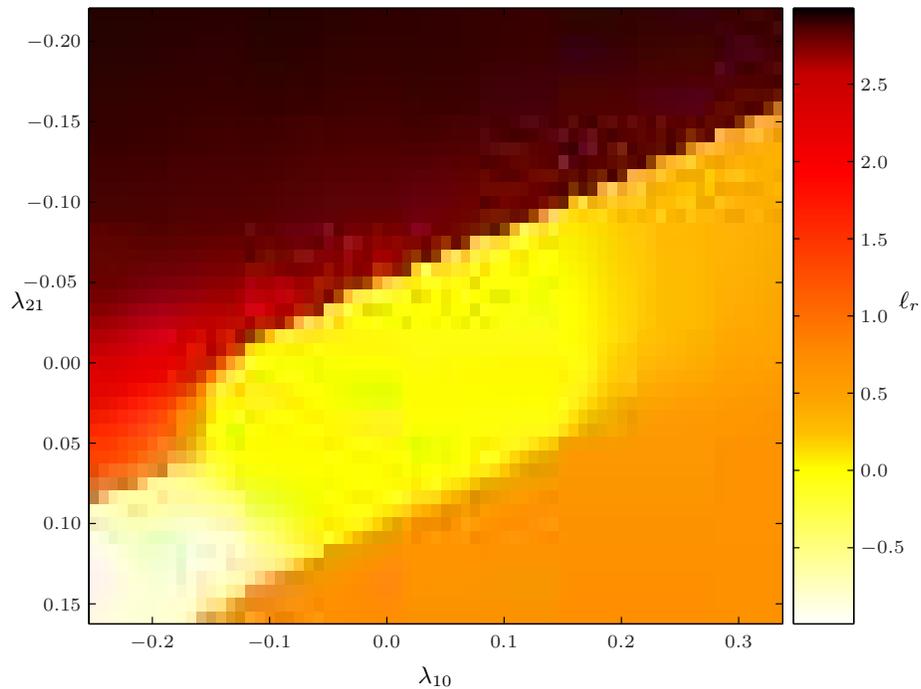}
\caption{\label{fig:phaseDiagMf} Phase diagram of the model with two couplings
as obtained via
MF approximation. The figure shows a contour plot of the ground
state expectation value  of $\ell_r$ in the
$\lambda_{10}$--$\lambda_{21}$ plane. The phase transition between
the two ferromagnetic phases (F-AC) is clearly visible in the
lower left part. Note that $\ell_r$ even discriminates between
symmetric and anti-ferromagnetic phases (S-AF) as can be seen in
the lower right part of the figure. A heuristic explanation of
this phenomenon is given in the main text.}
\end{figure*}

\begin{figure*}[!p]
\includegraphics{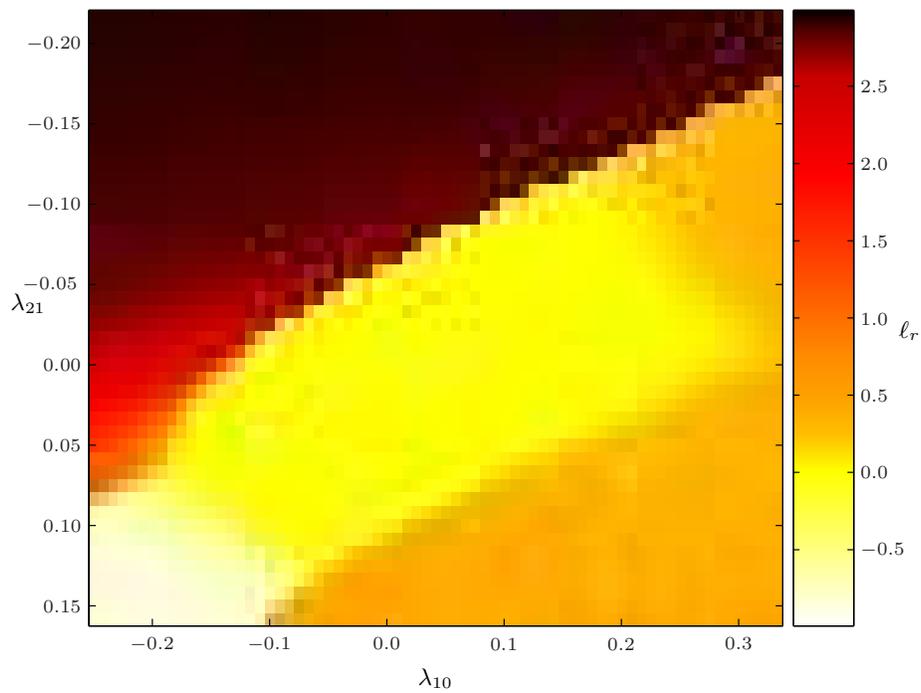}
\caption{\label{fig:phaseDiagMc8} Same phase diagram as in
Fig.~\protect\ref{fig:phaseDiagMf}, this time obtained via Monte
Carlo simulations on an $8^3$-lattice. Like in the minimal model MF
and Monte Carlo results agree quantitatively within an accuracy of
10\% and less. The distinction between S and AC will become more apparent in
Fig.\ref{fig:phaseStructure}.}
\end{figure*}

Our main results for the phase diagram are shown in
Fig.~\ref{fig:phaseDiagMf} obtained by the modified MF approximation
explained earlier and
Fig.~\ref{fig:phaseDiagMc8} which displays the Monte Carlo data and
hence constitutes the most faithful representation of the phase
structure.


The Monte Carlo simulations were carried out on a $8^3$-lattice. The
number of sweeps was chosen such as to reduce the
jackknife error in the estimate of $\ell_r$ below $0.1$. The
independence of our samples was ensured by demanding that the
autocorrelation time $\tau_{\ell_r}$ associated with the observable
$\ell_r$ was less then one percent of the total number of sweeps. As
a result our simulations included at least $4 \times 10^4$ sweeps
far away from the critical regions and more than $10^6$ sweeps in
their vicinity.

It is reassuring to note that the qualitative phase diagram
Fig.~\ref{fig:CLPHASES} predicted from energy-entropy arguments is
quantitatively confirmed by both Figs. \ref{fig:phaseDiagMf} and
\ref{fig:phaseDiagMc8}. Comparing the latter in some detail it is
once more remarkable how good the MF approximation works. Within
large regions of parameter space it agrees with the `real' data
within 10\% or less. Interestingly, the observable $\ell_r$ also
seems to be sensitive to the AF phase (see lower right part of Figs.
\ref{fig:phaseDiagMf} and \ref{fig:phaseDiagMc8}). However, any
further discrimination between phases F and AF by means of $\ell_r$
is impossible. To lift this degeneracy one clearly needs the AF
order parameter $\mathit{m}$ as an additional input.

It remains to be discussed why $\ell_r$ is sensitive at all to AF
ordering. In what follows we will provide a heuristic answer in the
context of the 3-state Potts model. By definition, all spins are
(anti)aligned in the (anti)ferromagnetic ground state.  In the
$\mathbb{Z}(2)$ symmetric Ising model with only two spin states each
possible ground state is two-fold degenerate, independent of the
particular ordering (F or AF). The counting of degeneracies,
however, is totally different for systems with more spin states
(hence higher symmetry). In the AF phase the additional freedom of
choice between two (or more) states anti-aligned with a given one
leads to an enormous degeneracy of the ground state \textit{in
energy}. As a consequence, \textit{entropy} will be the sole judge
deciding what is to be observed in a measurement. For the
$\mathbb{Z}(3)$ Polyakov loop model both MF approximation and the
Monte Carlo simulations tell us that the most probable ground state
corresponds to a preferred direction for the Polyakov loop on one of
the two sub-lattices and an equal distribution of the two remaining
directions on the other sub-lattice. Although we do not have an
analytical justification for this statement the numerical evidence
is compelling. Based on the latter, the sensitivity of $\ell_r$ to
AF ordering can be explained by a net expectation value which for
the case of the 3-state Potts model is easily computed as
\begin{equation}
  \ell_r = 0.5 \, z_1 + 0.25 \,  z_2 + 0.25 \, z_3 \neq 0 \; .
\end{equation}
\begin{figure}[t]
 \includegraphics{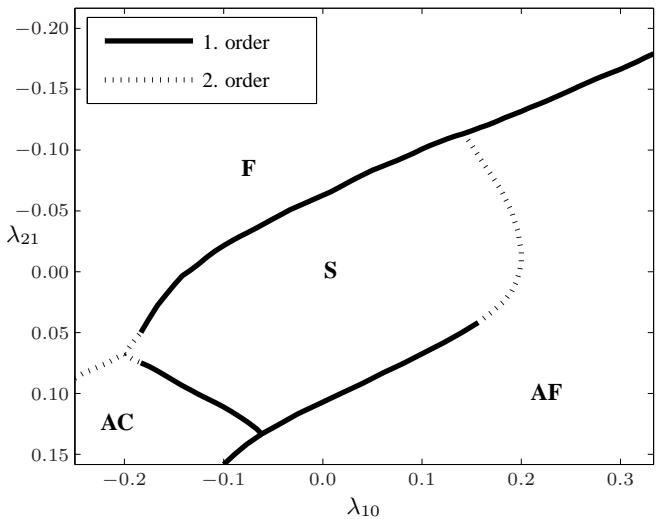}
 \caption{\label{fig:phaseStructure} Phase boundaries and orders of
 transitions as obtained via Monte Carlo simulation on a
 $8^3$-lattice. We observe a mixture of both first and second order
 transitions depending on the particular values for the couplings.
 The symmetric phase is enclosed by ordered phases as already
 expected from the discussion of Section~\ref{sec:QCA}. For further
 details on the simulation see the main text.}
\end{figure}%
\begin{figure}[t]
 \includegraphics{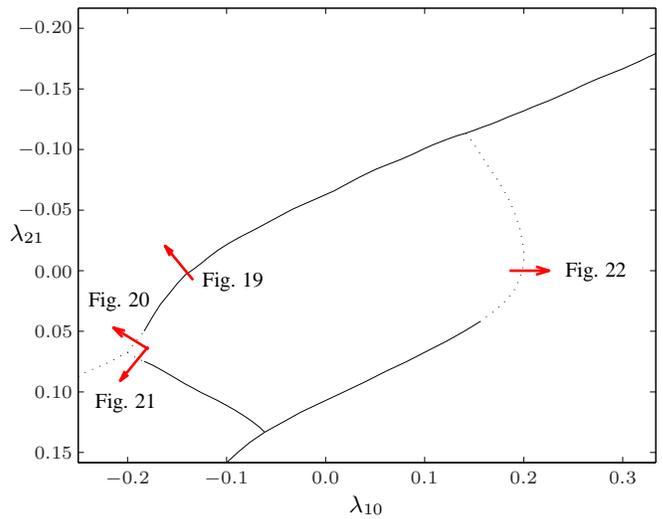}
 \caption{\label{fig:histoPositions} Phase diagram of
 Fig.~\ref{fig:phaseStructure} with oriented trajectories (marked by
 arrows) used for the histograms of Figs.~\ref{fig:histoOne} to
 \ref{fig:histoEight}. The curves are directed from the symmetric
 to the broken phases intersecting the critical lines vertically.
 We have chosen a representative subset from a total of twenty
 such curves which were analysed to determine the order of the
 transitions.}
\end{figure}

We conclude this discussion with an overview of the resulting
phases, their boundaries and the order of the transitions in
between presented in Fig.~\ref{fig:phaseStructure}. Our reasoning
here is based on additional measurements carried out along
parametrised curves $\boldsymbol\lambda = \boldsymbol \lambda(s)$
as depicted in Fig.~\ref{fig:histoPositions}. These simulations
were exclusively focused on the order of the phase transition
measuring the histograms of the observables $L$ and $M$. As before
all measurements were carried out on $8^3$-lattices each
trajectory $\boldsymbol\lambda(s)$ being sampled with twenty
points. For every such point $\boldsymbol \lambda_s$ $10^6$ Monte
Carlo sweeps were performed. To improve the statistics of the
histograms we made use of all previously discussed symmetries
($\mathbb{Z}(3)$, complex conjugation and exchange of even and odd
sub-lattice) by binning $L$ or $M$ together with their centre
images $z L$, $z^2 L$,  $z M$, $z^2 M$ and their complex
conjugates. This amounts to using each measured value of $L$ six
times and of $M$ even twelve times. In total we have recorded
twenty such runs four of which are depicted in
Fig.~\ref{fig:histoPositions}.

\begin{figure*}[!p]
 \includegraphics{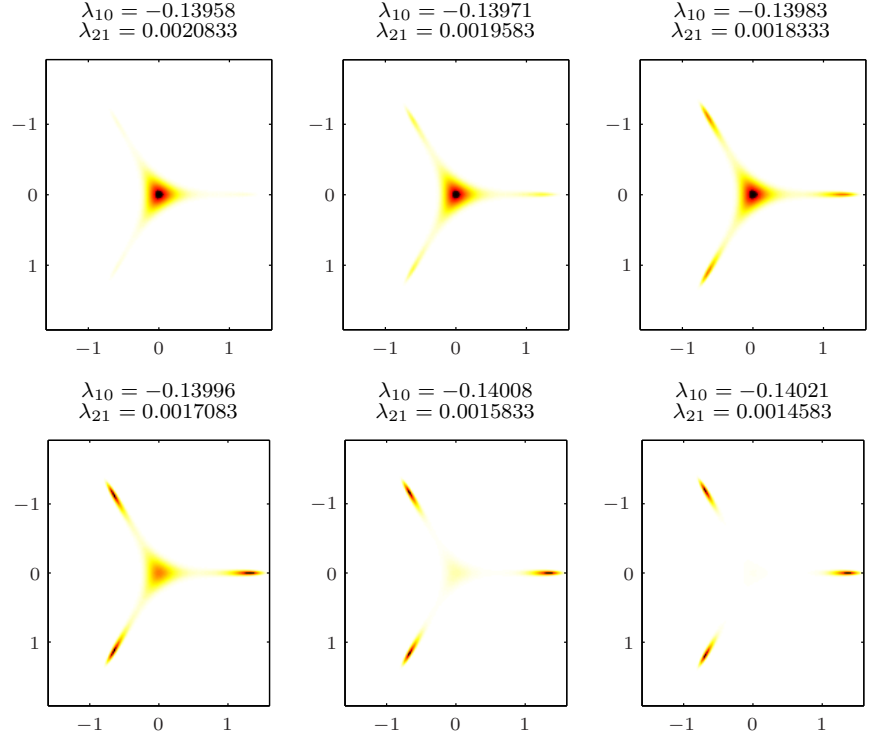}
 \caption{\label{fig:histoOne} Histogram of the observable $L$
 in the complex plane at a first order
 ferromagnetic phase transition.}
\end{figure*}

\begin{figure*}[!p]
 \includegraphics{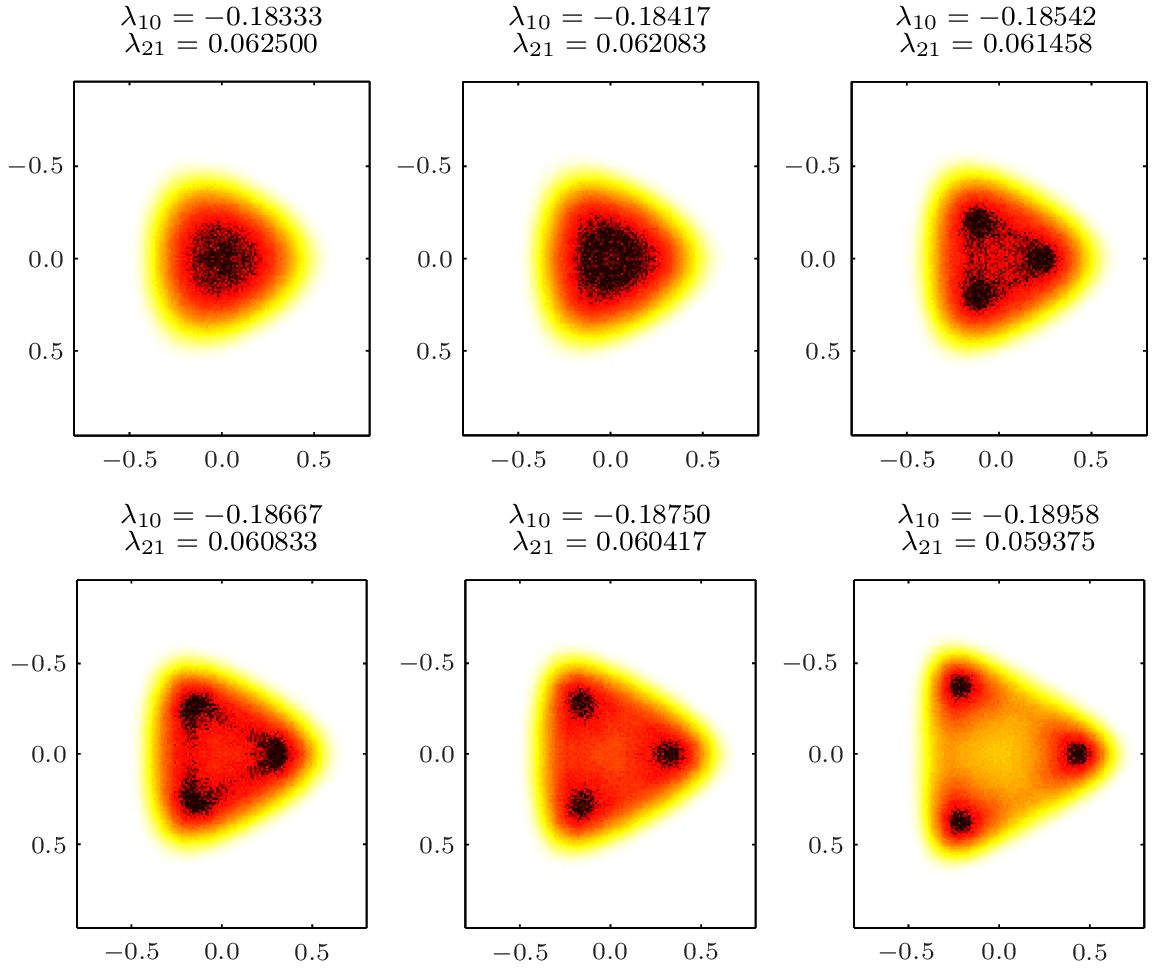}
 \caption{\label{fig:histoTwo} Histogram of the observable $L$ in
 the complex plane at a second order ferromagnetic phase
 transition.}
\end{figure*}

\begin{figure*}[!p]
 \includegraphics{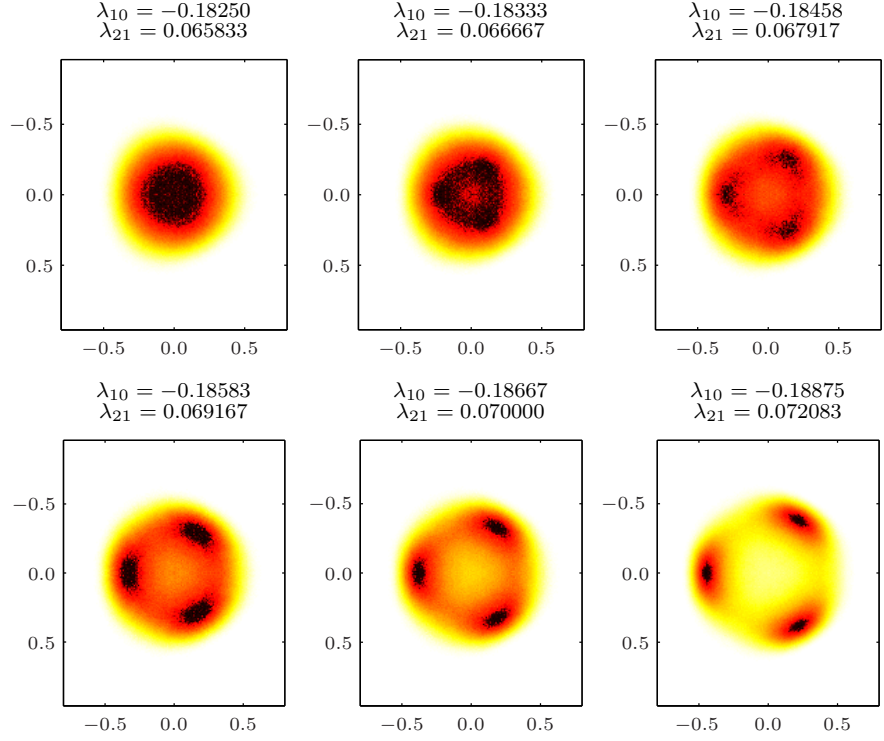}
 \caption{\label{fig:histoThree} Histogram of the observable $L$
 in the complex plane at a second order anti-centre phase
 transition.}
\end{figure*}

\begin{figure*}[!p]
 \includegraphics{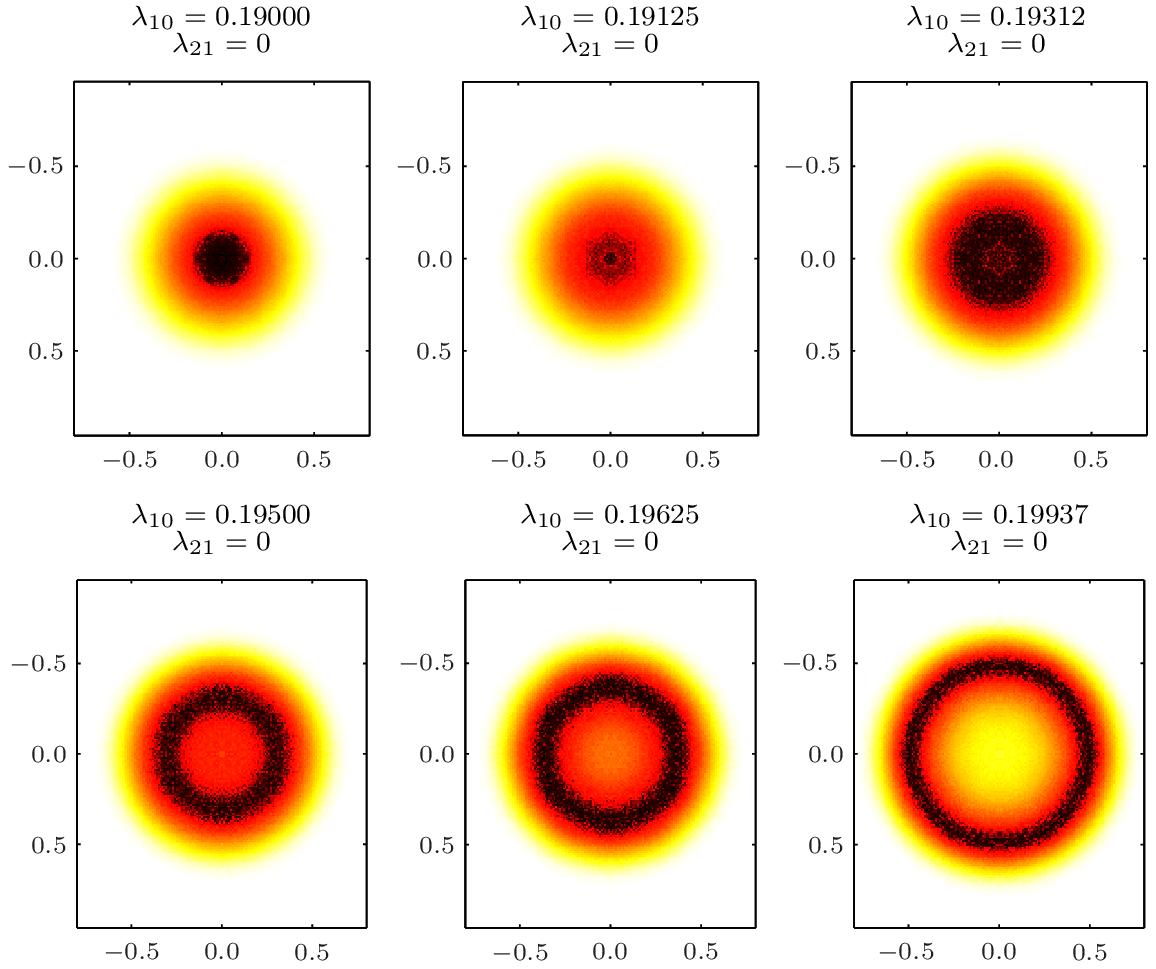}
 \caption{\label{fig:histoEight} Histogram of the observable $M$
 in the complex plane at a second order anti-ferromagnetic phase
 transition.}
\end{figure*}

To illuminate the order of the transitions corresponding to the
four directed line crossings displayed in
Fig.~\ref{fig:histoPositions} we present six (out of twenty)
histograms in Figs.~\ref{fig:histoOne} to \ref{fig:histoEight}
which display the distribution of the observable  $L$ respectively $M$ in the
complex plane.


The first figure in this series, Fig.~\ref{fig:histoOne},
corresponds to the ferromagnetic transition, S-F.  The histograms
displayed may be viewed as a `movie sequence' starting out in the
symmetric phase with the distribution $\rho(L)$ located at the
origin. Recalling the relation between probability distribution
$\rho(L)$ and the constraint effective potential
\cite{O'Raifeartaigh:1986hi}, $U(L) \propto \exp(-\rho(L))$, this
situation corresponds to a unique minimum of $U(L)$ at $L=0$. As
the couplings change three further maxima -- which are
$\mathbb{Z}(3)$ copies of each other -- arise in addition to the
one at the origin. Hence, we observe coexistence of ordered and
disordered phases. The new maxima are separated from the original
unique maximum at the origin by a finite amount. The associated
discontinuity together with coexistence clearly shows that the
transition is first order.

In contrast to this the situation depicted in
Fig.~\ref{fig:histoTwo} is quite different. First of all the
distribution is much more delocalised as compared to
Fig.~\ref{fig:histoOne} implying a much flatter effective
potential. Moreover,  as soon as the maxima of the broken
(ordered) phase emerge the maximum at the origin has dissolved.
Hence, there is no coexistence of different phases as in the
previous case but rather a continuous appearance of new maxima,
branching away from the origin towards the corners of the
fundamental domain. These features characterise a second order
phase transition.

Analogous behaviour can be observed in Fig.~\ref{fig:histoThree}
with the maxima now moving in roughly the opposite directions
indicating a transition to the AC phase. Still different is the
S-AF transition of Fig.~\ref{fig:histoEight}. By the same
reasoning as given above this transition is of second order as
well. However, the probability function emerging in the broken
phase does not display the usual three-fold symmetry we have
observed for the previous transitions. Although the figures
seemingly look rotationally invariant one actually finds a
six-fold degeneracy rather than a continuous one. The symmetry
enhancement by a factor of two has a simple explanation. Note that
Fig.~\ref{fig:histoEight} shows the probability distribution of
the AF order parameter $M$ which is defined in terms of even and
odd sub-lattices. Interchanging the latter accounts for the
additional factor of two.

Let us conclude this section with a short remark concerning
Figs.~\ref{fig:histoTwo} and \ref{fig:histoThree}. We have seen
that our numerical data seem to indicate the possibility of
second order transitions from the disordered phase S to F or AC
ordering. This is at variance with the folklore claiming the
absence of a $\mathbb{Z}(3)$ universality class in $d=3$, see
e.g.\ \cite{Holland:2003kg}. It is not excluded that the observed
second order characteristics near the tricritical point (S-F-AC) derive from
artifacts due to our algorithms. At the moment, however, we cannot make a
definite
statement about this issue. Obviously, the puzzle deserves further
investigation in the future.

\section{Summary and Outlook}
\label{sec:SO}

As outlined in the introduction the original motivation for
Polyakov loop models lies in their status as effective theories
for finite-temperature gluodynamics. However, our investigation of
the statistical mechanics involved should have made it clear that
they are interesting in their own right. In support of this
statement we mention the 3-fold $\mathbb{Z}(3)$ symmetry shared
with the 3-state Potts model, the nontrivial complex target space
and the enormously rich phase structure resulting thereof.

As we have seen, the latter can be understood qualitatively by
simple energy-entropy considerations which predict a symmetric
phase S close to the origin in coupling space `surrounded'  by
broken phases. These can exhibit ferromagnetic (F),
anti-ferromagnetic (AF) or anti-centre (AC) ordering. This picture
has been confirmed quantitatively both within a mean-field (MF)
approximation and by extensive Monte Carlo simulations. Both
methods find a first order transition for S-F and a second order
one for S-AF. In order to capture the details of the phase diagram
a total number of 8000 simulations (corresponding to roughly 3000
hours of CPU time) was required. Naturally, this has added further
refinements to our classical and MF analysis like precise values
for critical couplings and exponents.

The agreement between MF and Monte Carlo result is much better
than naively expected for a statistical model in $d=3$. The S-F
critical couplings agree to within 1\%, the discontinuity $\Delta
\ell$ within 10\% and the S-AF critical coupling within 20\%. The
precise agreement for the S-F transition suggests that the point
where the S-F and S-AC critical lines merge is actually a
\textit{tricritical} point where two first order transitions merge
into a second order one. It is known that the upper critical
dimension for a tricritical point is three \cite{PhysRevB.31.4701,Lawrie:1984}
so that in $d=3$ the MF approximation actually is exact apart from
logarithmic corrections \cite{Landau:1986}. This observation is
corroborated by the fact that the Polyakov loop model with two
couplings is somewhat similar to the spin-one Blume-Capel
\cite{Blume:1966,Capel:1966} model which apart from an Ising term
has an addition $K \sum_i s_i^2$, $s_i = -1, 0, 1$, with 3 spin
states. The model is known to have a tricritical point and is
closely related to the 3-state Potts model \cite{Landau:1981}. The
second order phase transition line is characterised by 3$d$ Ising
critical exponents. Obviously, one should test the analogous
exponents for our AC-F transition at the very left of our phase
diagrams.

One may also address the spatial localisation of the phases. On
the 3$d$ lattice the coexisting phases should be separated by
interfaces the tension of which can be measured (see e.g.\
\cite{Holland:2000uj} and references therein).

In the end one is of course interested in matching the effective
couplings to the underlying microscopic theory, i.e. \
finite-temperature Yang-Mills. The resulting curve
$\lambda_{IJ}(\beta)$ in the space of effective couplings
should stay in the symmetric and ferromagentic phases.
We have already solved the matching-problem
for the (simpler) case of $SU(2)$ using inverse Monte Carlo
methods based on Schwinger-Dyson equations. For $SU(3)$ this is
considerably more difficult (at least technically) due to the
increased complexity of the Haar measure. The problem is actually
similar to one encountered in strong-coupling expansions where
$SU(2)$ group integrals can be performed analytically which is no
longer true for $SU(3)$. Nevertheless, we have been able to derive
the relevant Schwinger-Dyson equations and hope to report on their
applications in the near future.

\section{Acknowledgements}
\label{sec:ACK} We thank Guy Buss for his contributions to the
strong coupling expansion and Georg Bergner, Wolfhard Janke
and Erhard Seiler for valuable hints. TK
acknowledges support by the Konrad-Adenauer-Stiftung e.V., CW by
the Studienstiftung des deutschen Volkes and TH by the Plymouth
Particle Theory Group.


\begin{thebibliography}{47}
\expandafter\ifx\csname natexlab\endcsname\relax\def\natexlab#1{#1}\fi
\expandafter\ifx\csname bibnamefont\endcsname\relax
  \def\bibnamefont#1{#1}\fi
\expandafter\ifx\csname bibfnamefont\endcsname\relax
  \def\bibfnamefont#1{#1}\fi
\expandafter\ifx\csname citenamefont\endcsname\relax
  \def\citenamefont#1{#1}\fi
\expandafter\ifx\csname url\endcsname\relax
  \def\url#1{\texttt{#1}}\fi
\expandafter\ifx\csname urlprefix\endcsname\relax\def\urlprefix{URL }\fi
\providecommand{\bibinfo}[2]{#2}
\providecommand{\eprint}[2][]{\url{#2}}

\bibitem[{\citenamefont{Yaffe and Svetitsky}(1982)}]{Yaffe:1982qf}
\bibinfo{author}{\bibfnamefont{L.~G.} \bibnamefont{Yaffe}} \bibnamefont{and}
  \bibinfo{author}{\bibfnamefont{B.}~\bibnamefont{Svetitsky}},
  \bibinfo{journal}{Phys. Rev.} \textbf{\bibinfo{volume}{D26}},
  \bibinfo{pages}{963} (\bibinfo{year}{1982}).

\bibitem[{\citenamefont{Svetitsky and Yaffe}(1982)}]{Svetitsky:1982gs}
\bibinfo{author}{\bibfnamefont{B.}~\bibnamefont{Svetitsky}} \bibnamefont{and}
  \bibinfo{author}{\bibfnamefont{L.~G.} \bibnamefont{Yaffe}},
  \bibinfo{journal}{Nucl. Phys.} \textbf{\bibinfo{volume}{B210}},
  \bibinfo{pages}{423} (\bibinfo{year}{1982}).

\bibitem[{\citenamefont{Engels et~al.}(1990)\citenamefont{Engels, Fingberg, and
  Weber}}]{Engels:1989fz}
\bibinfo{author}{\bibfnamefont{J.}~\bibnamefont{Engels}},
  \bibinfo{author}{\bibfnamefont{J.}~\bibnamefont{Fingberg}}, \bibnamefont{and}
  \bibinfo{author}{\bibfnamefont{M.}~\bibnamefont{Weber}},
  \bibinfo{journal}{Nucl. Phys.} \textbf{\bibinfo{volume}{B332}},
  \bibinfo{pages}{737} (\bibinfo{year}{1990}).

\bibitem[{\citenamefont{Engels et~al.}(1992)\citenamefont{Engels, Fingberg, and
  Miller}}]{Engels:1992fs}
\bibinfo{author}{\bibfnamefont{J.}~\bibnamefont{Engels}},
  \bibinfo{author}{\bibfnamefont{J.}~\bibnamefont{Fingberg}}, \bibnamefont{and}
  \bibinfo{author}{\bibfnamefont{D.~E.} \bibnamefont{Miller}},
  \bibinfo{journal}{Nucl. Phys.} \textbf{\bibinfo{volume}{B387}},
  \bibinfo{pages}{501} (\bibinfo{year}{1992}).

\bibitem[{\citenamefont{Holland
  et~al.}(2004{\natexlab{a}})\citenamefont{Holland, Pepe, and
  Wiese}}]{Holland:2003kg}
\bibinfo{author}{\bibfnamefont{K.}~\bibnamefont{Holland}},
  \bibinfo{author}{\bibfnamefont{M.}~\bibnamefont{Pepe}}, \bibnamefont{and}
  \bibinfo{author}{\bibfnamefont{U.~J.} \bibnamefont{Wiese}},
  \bibinfo{journal}{Nucl. Phys.} \textbf{\bibinfo{volume}{B694}},
  \bibinfo{pages}{35} (\bibinfo{year}{2004}{\natexlab{a}}),
  \eprint{hep-lat/0312022}.

\bibitem[{\citenamefont{Celik et~al.}(1983)\citenamefont{Celik, Engels, and
  Satz}}]{Celik:1983wz}
\bibinfo{author}{\bibfnamefont{T.}~\bibnamefont{Celik}},
  \bibinfo{author}{\bibfnamefont{J.}~\bibnamefont{Engels}}, \bibnamefont{and}
  \bibinfo{author}{\bibfnamefont{H.}~\bibnamefont{Satz}},
  \bibinfo{journal}{Phys. Lett.} \textbf{\bibinfo{volume}{B125}},
  \bibinfo{pages}{411} (\bibinfo{year}{1983}).

\bibitem[{\citenamefont{Kogut et~al.}(1983)}]{Kogut:1982rt}
\bibinfo{author}{\bibfnamefont{J.~B.} \bibnamefont{Kogut}}
  \bibnamefont{et~al.}, \bibinfo{journal}{Phys. Rev. Lett.}
  \textbf{\bibinfo{volume}{50}}, \bibinfo{pages}{393} (\bibinfo{year}{1983}).

\bibitem[{\citenamefont{Gottlieb et~al.}(1985)}]{Gottlieb:1985ug}
\bibinfo{author}{\bibfnamefont{S.~A.} \bibnamefont{Gottlieb}}
  \bibnamefont{et~al.}, \bibinfo{journal}{Phys. Rev. Lett.}
  \textbf{\bibinfo{volume}{55}}, \bibinfo{pages}{1958} (\bibinfo{year}{1985}).

\bibitem[{\citenamefont{Brown et~al.}(1988)\citenamefont{Brown, Christ, Deng,
  Gao, and Woch}}]{Brown:1988qe}
\bibinfo{author}{\bibfnamefont{F.~R.} \bibnamefont{Brown}},
  \bibinfo{author}{\bibfnamefont{N.~H.} \bibnamefont{Christ}},
  \bibinfo{author}{\bibfnamefont{Y.~F.} \bibnamefont{Deng}},
  \bibinfo{author}{\bibfnamefont{M.~S.} \bibnamefont{Gao}}, \bibnamefont{and}
  \bibinfo{author}{\bibfnamefont{T.~J.} \bibnamefont{Woch}},
  \bibinfo{journal}{Phys. Rev. Lett.} \textbf{\bibinfo{volume}{61}},
  \bibinfo{pages}{2058} (\bibinfo{year}{1988}).

\bibitem[{\citenamefont{Fukugita et~al.}(1989)\citenamefont{Fukugita, Okawa,
  and Ukawa}}]{Fukugita:1989yb}
\bibinfo{author}{\bibfnamefont{M.}~\bibnamefont{Fukugita}},
  \bibinfo{author}{\bibfnamefont{M.}~\bibnamefont{Okawa}}, \bibnamefont{and}
  \bibinfo{author}{\bibfnamefont{A.}~\bibnamefont{Ukawa}},
  \bibinfo{journal}{Phys. Rev. Lett.} \textbf{\bibinfo{volume}{63}},
  \bibinfo{pages}{1768} (\bibinfo{year}{1989}).

\bibitem[{\citenamefont{Alves et~al.}(1990)\citenamefont{Alves, Berg, and
  Sanielevici}}]{Alves:1990yq}
\bibinfo{author}{\bibfnamefont{N.~A.} \bibnamefont{Alves}},
  \bibinfo{author}{\bibfnamefont{B.~A.} \bibnamefont{Berg}}, \bibnamefont{and}
  \bibinfo{author}{\bibfnamefont{S.}~\bibnamefont{Sanielevici}},
  \bibinfo{journal}{Phys. Rev. Lett.} \textbf{\bibinfo{volume}{64}},
  \bibinfo{pages}{3107} (\bibinfo{year}{1990}).

\bibitem[{\citenamefont{Pisarski}(2000)}]{Pisarski:2000eq}
\bibinfo{author}{\bibfnamefont{R.~D.} \bibnamefont{Pisarski}},
  \bibinfo{journal}{Phys. Rev.} \textbf{\bibinfo{volume}{D62}},
  \bibinfo{pages}{111501} (\bibinfo{year}{2000}), \eprint{hep-ph/0006205}.

\bibitem[{\citenamefont{Pisarski}(2002)}]{Pisarski:2001pe}
\bibinfo{author}{\bibfnamefont{R.~D.} \bibnamefont{Pisarski}},
  \bibinfo{journal}{Nucl. Phys.} \textbf{\bibinfo{volume}{A702}},
  \bibinfo{pages}{151} (\bibinfo{year}{2002}), \eprint{hep-ph/0112037}.

\bibitem[{\citenamefont{Dumitru et~al.}(2004)\citenamefont{Dumitru, Hatta,
  Lenaghan, Orginos, and Pisarski}}]{Dumitru:2003hp}
\bibinfo{author}{\bibfnamefont{A.}~\bibnamefont{Dumitru}},
  \bibinfo{author}{\bibfnamefont{Y.}~\bibnamefont{Hatta}},
  \bibinfo{author}{\bibfnamefont{J.}~\bibnamefont{Lenaghan}},
  \bibinfo{author}{\bibfnamefont{K.}~\bibnamefont{Orginos}}, \bibnamefont{and}
  \bibinfo{author}{\bibfnamefont{R.~D.} \bibnamefont{Pisarski}},
  \bibinfo{journal}{Phys. Rev.} \textbf{\bibinfo{volume}{D70}},
  \bibinfo{pages}{034511} (\bibinfo{year}{2004}), \eprint{hep-th/0311223}.

\bibitem[{\citenamefont{Holland and Wiese}(2000)}]{Holland:2000uj}
\bibinfo{author}{\bibfnamefont{K.}~\bibnamefont{Holland}} \bibnamefont{and}
  \bibinfo{author}{\bibfnamefont{U.-J.} \bibnamefont{Wiese}}
  (\bibinfo{year}{2000}), \bibinfo{note}{in \textsl{At the Frontier of Particle
  Physics --- Handbook of QCD}, M.~Shifman, ed., World Scientific, Singapore,
  2001}, \eprint{hep-ph/0011193}.

\bibitem[{\citenamefont{Holland
  et~al.}(2004{\natexlab{b}})\citenamefont{Holland, Pepe, and
  Wiese}}]{Holland:2003mc}
\bibinfo{author}{\bibfnamefont{K.}~\bibnamefont{Holland}},
  \bibinfo{author}{\bibfnamefont{M.}~\bibnamefont{Pepe}}, \bibnamefont{and}
  \bibinfo{author}{\bibfnamefont{U.~J.} \bibnamefont{Wiese}},
  \bibinfo{journal}{Nucl. Phys. Proc. Suppl.} \textbf{\bibinfo{volume}{129}},
  \bibinfo{pages}{712} (\bibinfo{year}{2004}{\natexlab{b}}),
  \eprint{hep-lat/0309062}.

\bibitem[{\citenamefont{Holland et~al.}(2003)\citenamefont{Holland, Minkowski,
  Pepe, and Wiese}}]{Holland:2003jy}
\bibinfo{author}{\bibfnamefont{K.}~\bibnamefont{Holland}},
  \bibinfo{author}{\bibfnamefont{P.}~\bibnamefont{Minkowski}},
  \bibinfo{author}{\bibfnamefont{M.}~\bibnamefont{Pepe}}, \bibnamefont{and}
  \bibinfo{author}{\bibfnamefont{U.~J.} \bibnamefont{Wiese}},
  \bibinfo{journal}{Nucl. Phys.} \textbf{\bibinfo{volume}{B668}},
  \bibinfo{pages}{207} (\bibinfo{year}{2003}), \eprint{hep-lat/0302023}.

\bibitem[{\citenamefont{Borgs and Seiler}(1983)}]{Borgs:1983yk}
\bibinfo{author}{\bibfnamefont{C.}~\bibnamefont{Borgs}} \bibnamefont{and}
  \bibinfo{author}{\bibfnamefont{E.}~\bibnamefont{Seiler}},
  \bibinfo{journal}{Commun. Math. Phys.} \textbf{\bibinfo{volume}{91}},
  \bibinfo{pages}{329} (\bibinfo{year}{1983}).

\bibitem[{\citenamefont{Ford et~al.}(1999)\citenamefont{Ford, Tok, and
  Wipf}}]{Ford:1998mq}
\bibinfo{author}{\bibfnamefont{C.}~\bibnamefont{Ford}},
  \bibinfo{author}{\bibfnamefont{T.}~\bibnamefont{Tok}}, \bibnamefont{and}
  \bibinfo{author}{\bibfnamefont{A.}~\bibnamefont{Wipf}},
  \bibinfo{journal}{Phys. Lett.} \textbf{\bibinfo{volume}{B456}},
  \bibinfo{pages}{155} (\bibinfo{year}{1999}), \eprint{hep-th/9811248}.

\bibitem[{\citenamefont{Meisinger et~al.}(2002)\citenamefont{Meisinger, Miller,
  and Ogilvie}}]{Meisinger:2001cq}
\bibinfo{author}{\bibfnamefont{P.~N.} \bibnamefont{Meisinger}},
  \bibinfo{author}{\bibfnamefont{T.~R.} \bibnamefont{Miller}},
  \bibnamefont{and} \bibinfo{author}{\bibfnamefont{M.~C.}
  \bibnamefont{Ogilvie}}, \bibinfo{journal}{Phys. Rev.}
  \textbf{\bibinfo{volume}{D65}}, \bibinfo{pages}{034009}
  (\bibinfo{year}{2002}), \eprint{hep-ph/0108009}.

\bibitem[{\citenamefont{Dittmann et~al.}(2004)\citenamefont{Dittmann, Heinzl,
  and Wipf}}]{Dittmann:2003qt}
\bibinfo{author}{\bibfnamefont{L.}~\bibnamefont{Dittmann}},
  \bibinfo{author}{\bibfnamefont{T.}~\bibnamefont{Heinzl}}, \bibnamefont{and}
  \bibinfo{author}{\bibfnamefont{A.}~\bibnamefont{Wipf}},
  \bibinfo{journal}{JHEP} \textbf{\bibinfo{volume}{06}}, \bibinfo{pages}{005}
  (\bibinfo{year}{2004}), \eprint{hep-lat/0306032}.

\bibitem[{\citenamefont{Heinzl et~al.}(2005)\citenamefont{Heinzl, Kaestner, and
  Wipf}}]{Heinzl:2005xv}
\bibinfo{author}{\bibfnamefont{T.}~\bibnamefont{Heinzl}},
  \bibinfo{author}{\bibfnamefont{T.}~\bibnamefont{Kaestner}}, \bibnamefont{and}
  \bibinfo{author}{\bibfnamefont{A.}~\bibnamefont{Wipf}},
  \bibinfo{journal}{Phys. Rev.} \textbf{\bibinfo{volume}{D72}},
  \bibinfo{pages}{065005} (\bibinfo{year}{2005}), \eprint{hep-lat/0502013}.

\bibitem[{\citenamefont{Polonyi and Szlachanyi}(1982)}]{Polonyi:1982wz}
\bibinfo{author}{\bibfnamefont{J.}~\bibnamefont{Polonyi}} \bibnamefont{and}
  \bibinfo{author}{\bibfnamefont{K.}~\bibnamefont{Szlachanyi}},
  \bibinfo{journal}{Phys. Lett.} \textbf{\bibinfo{volume}{B110}},
  \bibinfo{pages}{395} (\bibinfo{year}{1982}).

\bibitem[{\citenamefont{Ogilvie}(1984)}]{Ogilvie:1983ss}
\bibinfo{author}{\bibfnamefont{M.}~\bibnamefont{Ogilvie}},
  \bibinfo{journal}{Phys. Rev. Lett.} \textbf{\bibinfo{volume}{52}},
  \bibinfo{pages}{1369} (\bibinfo{year}{1984}).

\bibitem[{\citenamefont{Billo et~al.}(1996)\citenamefont{Billo, Caselle,
  D'Adda, and Panzeri}}]{Billo:1996wv}
\bibinfo{author}{\bibfnamefont{M.}~\bibnamefont{Billo}},
  \bibinfo{author}{\bibfnamefont{M.}~\bibnamefont{Caselle}},
  \bibinfo{author}{\bibfnamefont{A.}~\bibnamefont{D'Adda}}, \bibnamefont{and}
  \bibinfo{author}{\bibfnamefont{S.}~\bibnamefont{Panzeri}},
  \bibinfo{journal}{Nucl. Phys.} \textbf{\bibinfo{volume}{B472}},
  \bibinfo{pages}{163} (\bibinfo{year}{1996}), \eprint{hep-lat/9601020}.

\bibitem[{\citenamefont{Svetitsky}(1986)}]{Svetitsky:1985ye}
\bibinfo{author}{\bibfnamefont{B.}~\bibnamefont{Svetitsky}},
  \bibinfo{journal}{Phys. Rept.} \textbf{\bibinfo{volume}{132}},
  \bibinfo{pages}{1} (\bibinfo{year}{1986}).

\bibitem[{\citenamefont{Buss}(2004)}]{Buss:2004}
\bibinfo{author}{\bibfnamefont{G.}~\bibnamefont{Buss}} (\bibinfo{year}{2004}),
  \bibinfo{note}{diploma thesis, Jena (in German)}.

\bibitem[{\citenamefont{Creutz}(1983)}]{Creutz:1984mg}
\bibinfo{author}{\bibfnamefont{M.}~\bibnamefont{Creutz}},
  \emph{\bibinfo{title}{Quarks, Gluons and Lattices}}, {C}ambridge {M}onographs
  on {M}athematical {P}hysics (\bibinfo{publisher}{{C}ambridge {U}niversity
  {P}ress}, \bibinfo{year}{1983}).

\bibitem[{\citenamefont{Montvay and M{\"u}nster}(1994)}]{Montvay:1994cy}
\bibinfo{author}{\bibfnamefont{I.}~\bibnamefont{Montvay}} \bibnamefont{and}
  \bibinfo{author}{\bibfnamefont{G.}~\bibnamefont{M{\"u}nster}},
  \emph{\bibinfo{title}{Quantum Fields on a Lattice}}, {C}ambridge {M}onographs
  on {M}athematical {P}hysics (\bibinfo{publisher}{{C}ambridge {U}niversity
  {P}ress}, \bibinfo{year}{1994}).

\bibitem[{\citenamefont{Jensen and Mouritsen}(1979)}]{KnakJensen:1979}
\bibinfo{author}{\bibfnamefont{S.~J.~K.} \bibnamefont{Jensen}}
  \bibnamefont{and} \bibinfo{author}{\bibfnamefont{O.~G.}
  \bibnamefont{Mouritsen}}, \bibinfo{journal}{Phys. Rev. Lett.}
  \textbf{\bibinfo{volume}{43}}, \bibinfo{pages}{1763} (\bibinfo{year}{1979}).

\bibitem[{\citenamefont{Kogut and Sinclair}(1982)}]{Kogut:1982ss}
\bibinfo{author}{\bibfnamefont{J.~B.} \bibnamefont{Kogut}} \bibnamefont{and}
  \bibinfo{author}{\bibfnamefont{D.}~\bibnamefont{Sinclair}},
  \bibinfo{journal}{Solid State Commun.} \textbf{\bibinfo{volume}{41}},
  \bibinfo{pages}{187} (\bibinfo{year}{1982}).

\bibitem[{\citenamefont{Wu}(1982)}]{Wu:1982ra}
\bibinfo{author}{\bibfnamefont{F.~Y.} \bibnamefont{Wu}}, \bibinfo{journal}{Rev.
  Mod. Phys.} \textbf{\bibinfo{volume}{54}}, \bibinfo{pages}{235}
  (\bibinfo{year}{1982}).

\bibitem[{\citenamefont{Potts}(1952)}]{Potts:1952}
\bibinfo{author}{\bibfnamefont{R.}~\bibnamefont{Potts}},
  \bibinfo{journal}{Proc. Camb. Phil. Soc.} \textbf{\bibinfo{volume}{48}},
  \bibinfo{pages}{106} (\bibinfo{year}{1952}).

\bibitem[{\citenamefont{Janke and Villanova}(1997)}]{Janke:1996qb}
\bibinfo{author}{\bibfnamefont{W.}~\bibnamefont{Janke}} \bibnamefont{and}
  \bibinfo{author}{\bibfnamefont{R.}~\bibnamefont{Villanova}},
  \bibinfo{journal}{Nucl. Phys.} \textbf{\bibinfo{volume}{B489}},
  \bibinfo{pages}{679} (\bibinfo{year}{1997}), \eprint{hep-lat/9612008}.

\bibitem[{\citenamefont{Gottlob and Hasenbusch}(1994)}]{Gottlob:1994ds}
\bibinfo{author}{\bibfnamefont{A.~P.} \bibnamefont{Gottlob}} \bibnamefont{and}
  \bibinfo{author}{\bibfnamefont{M.}~\bibnamefont{Hasenbusch}},
  \bibinfo{journal}{Physica} \textbf{\bibinfo{volume}{A210}},
  \bibinfo{pages}{217} (\bibinfo{year}{1994}).

\bibitem[{\citenamefont{Wang et~al.}(1990)\citenamefont{Wang, Swendsen, and
  Koteck{\'y}}}]{Wang:1990}
\bibinfo{author}{\bibfnamefont{J.-S.} \bibnamefont{Wang}},
  \bibinfo{author}{\bibfnamefont{R.~H.} \bibnamefont{Swendsen}},
  \bibnamefont{and}
  \bibinfo{author}{\bibfnamefont{R.}~\bibnamefont{Koteck{\'y}}},
  \bibinfo{journal}{Phys. Rev.} \textbf{\bibinfo{volume}{B42}},
  \bibinfo{pages}{2465} (\bibinfo{year}{1990}).

\bibitem[{\citenamefont{Berg}(2000)}]{Berg:1998nj}
\bibinfo{author}{\bibfnamefont{B.~A.} \bibnamefont{Berg}},
  \bibinfo{journal}{Fields Inst. Commun.} \textbf{\bibinfo{volume}{26}},
  \bibinfo{pages}{1} (\bibinfo{year}{2000}), \eprint{cond-mat/9909236}.

\bibitem[{\citenamefont{Wolff}(1989)}]{Wolff:1988uh}
\bibinfo{author}{\bibfnamefont{U.}~\bibnamefont{Wolff}},
  \bibinfo{journal}{Phys. Rev. Lett.} \textbf{\bibinfo{volume}{62}},
  \bibinfo{pages}{361} (\bibinfo{year}{1989}).

\bibitem[{\citenamefont{Swendsen and Wang}(1987)}]{Swendsen:1987ce}
\bibinfo{author}{\bibfnamefont{R.~H.} \bibnamefont{Swendsen}} \bibnamefont{and}
  \bibinfo{author}{\bibfnamefont{J.-S.} \bibnamefont{Wang}},
  \bibinfo{journal}{Phys. Rev. Lett.} \textbf{\bibinfo{volume}{58}},
  \bibinfo{pages}{86} (\bibinfo{year}{1987}).

\bibitem[{\citenamefont{Gottlob and Hasenbusch}(1993)}]{Gottlob:1993zd}
\bibinfo{author}{\bibfnamefont{A.~P.} \bibnamefont{Gottlob}} \bibnamefont{and}
  \bibinfo{author}{\bibfnamefont{M.}~\bibnamefont{Hasenbusch}}
  (\bibinfo{year}{1993}), \eprint{cond-mat/9305020}.

\bibitem[{\citenamefont{O'Raifeartaigh
  et~al.}(1986)\citenamefont{O'Raifeartaigh, Wipf, and
  Yoneyama}}]{O'Raifeartaigh:1986hi}
\bibinfo{author}{\bibfnamefont{L.}~\bibnamefont{O'Raifeartaigh}},
  \bibinfo{author}{\bibfnamefont{A.}~\bibnamefont{Wipf}}, \bibnamefont{and}
  \bibinfo{author}{\bibfnamefont{H.}~\bibnamefont{Yoneyama}},
  \bibinfo{journal}{Nucl. Phys.} \textbf{\bibinfo{volume}{B271}},
  \bibinfo{pages}{653} (\bibinfo{year}{1986}).

\bibitem[{\citenamefont{Lawrie and S.Sarbach}(1984)}]{Lawrie:1984}
\bibinfo{author}{\bibfnamefont{I.}~\bibnamefont{Lawrie}} \bibnamefont{and}
  \bibinfo{author}{\bibnamefont{S.Sarbach}}, pp. \bibinfo{pages}{2--155}
  (\bibinfo{year}{1984}), \bibinfo{note}{in: C. Domb and J.L. Lebowitz,
  editors, Phase Transitions and Critical Phenomena, Vol. 9, chapter 1, pages
  2-155. Academic Press, New York}.

\bibitem[{\citenamefont{Lipowsky and Seifert}(1985)}]{PhysRevB.31.4701}
\bibinfo{author}{\bibfnamefont{R.}~\bibnamefont{Lipowsky}} \bibnamefont{and}
  \bibinfo{author}{\bibfnamefont{U.}~\bibnamefont{Seifert}},
  \bibinfo{journal}{Phys. Rev. B} \textbf{\bibinfo{volume}{31}},
  \bibinfo{pages}{4701} (\bibinfo{year}{1985}).

\bibitem[{\citenamefont{Landau and Swendsen}(1986)}]{Landau:1986}
\bibinfo{author}{\bibfnamefont{D.}~\bibnamefont{Landau}} \bibnamefont{and}
  \bibinfo{author}{\bibfnamefont{R.}~\bibnamefont{Swendsen}},
  \bibinfo{journal}{Phys. Rev. B} \textbf{\bibinfo{volume}{33}},
  \bibinfo{pages}{7700} (\bibinfo{year}{1986}).

\bibitem[{\citenamefont{Blume}(1966)}]{Blume:1966}
\bibinfo{author}{\bibfnamefont{M.}~\bibnamefont{Blume}},
  \bibinfo{journal}{Phys. Rev.} \textbf{\bibinfo{volume}{141}},
  \bibinfo{pages}{517} (\bibinfo{year}{1966}).

\bibitem[{\citenamefont{Capel}(1966)}]{Capel:1966}
\bibinfo{author}{\bibfnamefont{H.}~\bibnamefont{Capel}},
  \bibinfo{journal}{Physica} \textbf{\bibinfo{volume}{32}},
  \bibinfo{pages}{966} (\bibinfo{year}{1966}).

\bibitem[{\citenamefont{Landau and Swendsen}(1991)}]{Landau:1981}
\bibinfo{author}{\bibfnamefont{D.}~\bibnamefont{Landau}} \bibnamefont{and}
  \bibinfo{author}{\bibfnamefont{R.}~\bibnamefont{Swendsen}},
  \bibinfo{journal}{Phys. Rev. Lett.} \textbf{\bibinfo{volume}{46}},
  \bibinfo{pages}{1437} (\bibinfo{year}{1991}).

\end{thebibliography}
\end{document}